\documentclass[pre,twocolumn,superscriptaddress,floatfix]{revtex4}
\usepackage{epsfig}
\usepackage{float}
\usepackage{subfigure}
\usepackage{graphicx}
\usepackage{amssymb}

\begin{document}

\title{Inverse melting in a two-dimensional off-lattice model}

\author{Ahmad M. Almudallal}
\affiliation{Department of Physics and Physical Oceanography,
Memorial University of Newfoundland, St. John's, NL, A1B 3X7, Canada}

\author{Sergey V. Buldyrev}
\affiliation{Department of Physics, Yeshiva University,
500W 185th Street, New York, NY 10033, USA}

\author{Ivan Saika-Voivod}
\affiliation{Department of Physics and Physical Oceanography,
Memorial University of Newfoundland, St. John's, NL, A1B 3X7, Canada}

\date{\today}

\begin{abstract}

We carry out computer simulations of a simple, two-dimensional off-lattice model that exhibits inverse melting.  
The monodisperse system comprises core-softened disks interacting through a repulsive square shoulder located inside an attractive square well.  
By systematically varying the potential parameters, we increase the pressure range over which the liquid freezes to a crystal 
upon isobaric heating. The effect is largely controlled by the extent of the shoulder.  
Despite occurring in two dimensions, the melting transition is first order and to a liquid, rather than to a hexatic or quasicrystal phase.
We also provide comment on a commonly employed correlation function used to determine the degree of translational ordering in a system.

\end{abstract}

\maketitle

\section{Introduction}

Inverse melting is the curious phenomenon in which a crystal melts upon isobaric cooling, or equivalently, a liquid freezes
upon heating.  Only a handful of systems exhibit this rare behavior~\cite{stillinger2}.  While the effect is inherently fascinating, 
recent theoretical work on DNA-coated colloids points to inverse melting as a way to overcome
kinetic trapping at low temperature $T$, thus providing alternate pathways in the synthesis of 
novel materials~\cite{angiolettiuberti,gang}.

Notable examples of materials exhibiting inverse melting are $^3$He~\cite{dobbs} and $^4$He~\cite{pair, wilks}, 
for which the liquid is stabilized at low $T$ by quantum mechanical effects, and polymers 
poly(4-methylpentene-1)~\cite{rastogi1, rastogi2, rastogi3, greer}
and syndiotactic polystyrene~\cite{corstjens}.
Motivated by He and polymers, Feeney and coworkers devised a model that successfully recovers inverse melting by coupling 
internal degrees of freedom of a particle with
interparticle interactions~\cite{feeney}.
A lattice model for which the ferromagnetic phase receives an energetic penalty but is given a higher degeneracy also recovers
inverse melting~\cite{schupper}.
Other cases of inverse melting and behavior similar to it are the nematic to smectic-A transition achieved upon heating the liquid crystal 
4-cyano-4$^\prime$-octyloxybiphenyl~\cite{cladis}; crystallization of micelles of triblock copolymer PEO-PPO-PEO upon heating, brought
about by an increase in effective packing fraction as $T$ increases~\cite{mortensen}; 
the multicomponent solution of  a-cyclodextrine, water, and 4-methylpyridine~\cite{plazanet,angelini1,angelini2}
in which hydrogen bond rearrangements play a role; Nb-Cr alloys, which again are multicomponent solutions;
and the melting
of the ordered vortex phase in a high-temperature superconductor~\cite{avraham,beidenkopf}.

The idea of inverse melting is also linked conceptually to the glass transition.  It was pointed out by Kauzmann~\cite{kauzmann}
that in many cases, the behavior of a liquid cooled progressively below its freezing $T$ extrapolates to the thermodynamically exotic
case of the liquid's entropy becoming lower than that of the crystal.  Before this point is reached, the glass transition, 
a kinetic phenomenon, intervenes, implicating entropy as a controlling factor in liquid dynamics.
Inverse melting, however, requires that the crystal have 
a higher entropy than the liquid's over a range of thermodynamic conditions, a conclusion reached upon considering slopes of
melting lines in the pressure ($P$)-$T$ plane~\cite{stillinger2}.  So while systems exhibiting inverse melting provide a counter-example to the
importance of excess entropy to dynamics, they do provide the intriguing case in which a crystal may be quenched into a kinetically trapped metastable state, an ordered version of a glass~\cite{tombari}.  Further, one may wish to explore the possible connection between inverse melting
and glassy dynamics achieved upon heating~\cite{vargas}. 

What would enhance the current body of work on inverse melting is a simple off-lattice model that exhibits the phenomenon.
Recently, we reported inverse melting for a double-step potential consisting of a square shoulder within a square well (SSSW),
shown in Fig.~\ref{potential}, while calculating the phase diagram for the model in two dimensions 
[Fig.~\ref{phase_diagram_original}]~\cite{ahmad2}.  
The difficulty is that the effect is very weak, and we did not provide direct evidence for the existence of the phenomenon to confirm
the Monte Carlo-based free energy calculations used to determine phase boundaries.

In our present study, we tune the parameters of the model in a systematic way in order to greatly expand the region in
the $P$-$T$ plane over which inverse melting takes place.  Having enlarged the effect, we probe it with complementary 
techniques, including event-driven molecular dynamics (EDMD) simulations, to confirm its existence. 
Since in two dimensions there is the possibility of continuous melting through a hexatic-type phase we further provide 
evidence that the transition is first order between a liquid and crystal. Further, a quasicrystal phase for a similar potential has been
reported~\cite{skibinsky}, but we do not see such a phase.

The SSSW potential we study here falls into the category of core-softend potentials introduced by 
Stell and Hemmer~\cite{hemmer,stell} as model systems exhibiting multiple fluid or iso-structural solid transitions and critical 
points~\cite{denton1, denton2}.
Such potentials were used to study liquid metals~\cite{mon1,selbert,levesque,kincaid,cummings,velasco}, for which
experimental evidence exists for novel critical behavior~\cite{voronel}.
Research into explaining the many anomalous properties of water~\cite{kumar,debenedetti, prielmeier, kell},
particularly through a hypothesized second critical point in the deeply metastable state~\cite{poole}, has also
drawn benefit from studies of core-softened 
potentials~\cite{cho,sadr1,sadr2, jagla1,jagla2,jagla3,sadr3,sergey,saija, franzese, mausbach,quigley, head1,head2,hus}.

The particular model (in two dimensions) we use here was introduced in Ref.~\cite{sadr3} and further 
studied in Ref.~\cite{sergey}.  The model parameters were originally chosen so that a low density triangular crystal (LDT) and
a higher density square crystal (S) would have the same energy.  The competition between these two structures
within the liquid gives rise to anomalous properties, e.g., a line of density maxima.  A liquid-liquid critical point
is not observed in this 2D system, perhaps because of lack of strong metastability of the liquid below the LDT and S
melting lines near the triple point~\cite{ahmad2}.
Given that the model exhibits several anomalies, including two crystals that are less dense than the melt~\cite{ahmad2},
it is perhaps fitting that it also exhibits inverse melting.

This paper is organized as follows.  
In Section II, we discuss the model and the free energy and computer simulation techniques used in carrying out this work. 
In Section III, we give our results, including how potential parameters affect the melting line of the S crystal,
an estimate of the surface tension between liquid and crystal at low $T$ coexistence, direct MD simulations showing both metastablity 
and nucleation of S and the liquid, as well as structural measures that provide evidence against the existence of a hexatic-type phase or 
quasicrystals near the point of inverse melting.
In Section IV we provide a discussion and our conclusions.

\section{Methods}

\subsection{Model and simulations}

\begin{figure}[ht]
\centering\includegraphics[clip=true, trim=0 0 0 0, height=6.0cm, width=8.0cm]{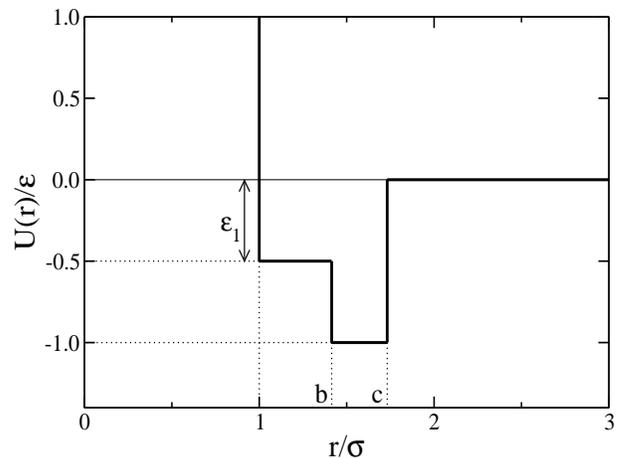}
\caption{Square-shoulder square-well potential with hard-core diameter $\sigma$ and bond energy $\epsilon$ as a function of particle separation $r$.
The original model parameters~\cite{sadr3} are soft-core distance
 $b=\sqrt{2}\sigma$, shoulder depth $\epsilon_1=\epsilon/2$ and limit of attraction $c=\sqrt{3}\sigma$.}
\label{potential}
\end{figure}

The SSSW interaction potential $U(r)$ that we consider in this study is a double-step potential consisting of 
a square shoulder and a square well as shown in Fig.~\ref{potential}. 
We study the potential in two dimensions, in which it describes disks with a hard-core diameter 
$\sigma$ followed by a square shoulder of interaction energy $-\epsilon_1$ for $\sigma < r < b$. 
The shoulder is followed by a square well of energy $-\epsilon$ for $b < r < c$. 
As in Ref.~\cite{sadr3}, we start with  potential parameters $\epsilon_1=\epsilon/2$,  
$b=\sqrt{2}\sigma$, and $c=\sqrt{3}\sigma$. 
The three parameters were originally assigned these values in order to bestow two crystals of different density, LDT and S, the 
same potential energy per particle of $-3\epsilon$, 
i.e., to create two energetically degenerate phases of well separated densities~\cite{sergey}. 
The idea behind this is to allow for distinct liquid states, one based on square packing and the other on the more 
open triangular lattice, in analogy to what is thought to be the case for water. 

\begin{figure}[ht]
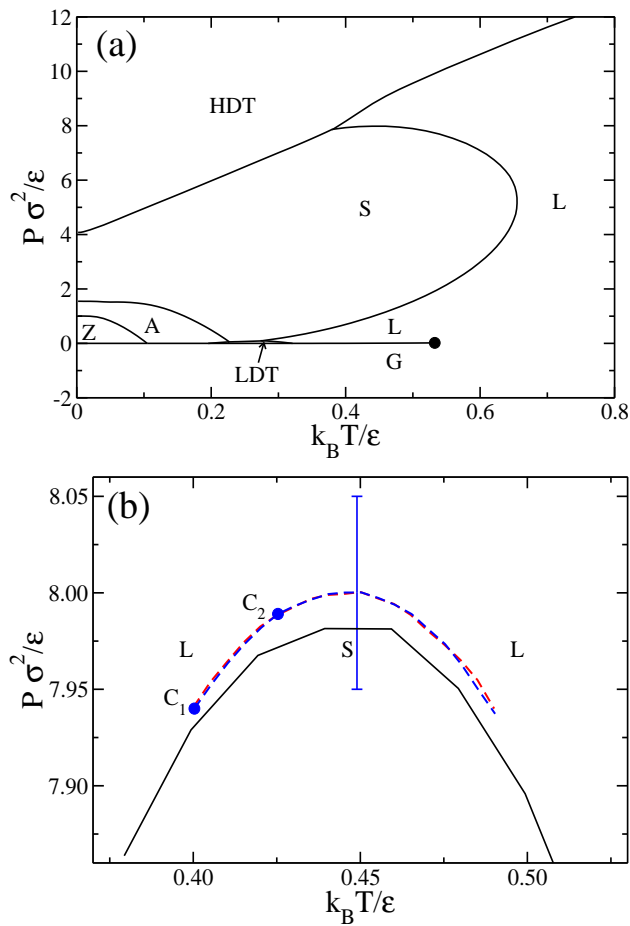

\subfigure{
\centering\includegraphics[clip=true, trim=0 0 0 0, height=6.0cm, width=8.25cm]{fig2a}
\label{phase_diagram_original}
}
\subfigure{
\label{inverse_melting_slope-dp}
\centering\includegraphics[clip=true, trim=0 0 0 0, height=6.0cm, width=8.25cm]{fig2b}
}
\caption{Panel (a) shows the phase diagram of the SSSW potential model with potential parameters $\epsilon_1=\epsilon/2$, $b=\sqrt{2}\sigma$ and $c=\sqrt{3}\sigma$ (adapted from Ref.~\cite{ahmad2}). 
Panel (b) is a close up of the maximum pressure of the L-S line. $C_1$ is the lower melting point along $P \sigma^2/\epsilon=7.94$,
from which Gibbs-Duhem integration is carried out to determine the coexistence line to the higher melting point at the same pressure (blue dashed curve). 
The red dashed coexistence curve results from integrating from the higher melting $T$ back to $C_1$.  
Both $C_1$ and $C_2$ are points on the coexistence line from which Hamiltonian Gibbs-Duhem integration 
is carried out to explore the effect of changing model parameters. }
\label{phase_diagram}
\end{figure}

In Ref~\cite{ahmad2}, we used various Monte Carlo simulation techniques to calculate the phase diagram 
for the same interaction potential over a wide range of temperature and pressure, as shown in Fig.~\ref{phase_diagram_original},
for the liquid (L), gas (G) and  five crystal phases:
the close-packed high-density triangular (HDT) crystal, LDT, S, and two low-$T$ crystals A and Z.
Apart from the case of the L-HDT transition at high $T$,
the methods used to calculate phase boundaries required metastability of the phases concerned, and therefore
provided evidence that the transitions are first order.

We also found that the S-L melting line exhibits a maximum temperature, 
as well as a maximum pressure that implied inverse melting over a very small range in pressure.   
We did not, however, provide any strong direct evidence that the model
exhibits inverse melting.
Our goal in the present study is to find  potential parameters $\epsilon_1$,
$b$ and $c$ that significantly increase the range of pressure over which inverse melting occurs, so that 
it can be observed more easily.

In this study, our calculations are based on free energy techniques that employ standard Metropolis MC simulations 
performed at constant number of particles $N$, $P$, and $T$, i.e., in the $NPT$ ensemble~\cite{frenkel1}.  
We simulate 1024 particles in a square box with periodic boundary conditions 
and we change the box size isotropically to maintain its square shape. 
To observe the liquid freeze after increasing $T$ and the crystal melt upon decreasing $T$ with an independent method, 
we carry out EDMD simulations~\cite{at,rapaport,adler,lubachevsky} of up to 65536 particles.

\subsection{Square crystal-liquid coexistence}

Although we calculated the S melting line for the SSSW model with its original parameters in Ref.~\cite{ahmad2} and
found good consistency between traces of the coexistence curve starting at independent initial coexistence points,
we wish to recalculate the curve since the inverse melting effect is so small.  Our present approach is to calculate the 
chemical potential for both S ($\mu_{\rm S}$) and L ($\mu_{\rm L}$) as a function of $T$ along $P \sigma^2/\epsilon=7.94$, a pressure at which
$\mu_{\rm L}(T)$ and $\mu_{\rm S}(T)$ should cross twice, since this pressure should be in the middle of the narrow inverse melting pressure range,
as shown in Fig.~\ref{inverse_melting_slope-dp}, and there should be two melting temperatures.

For the S crystal, we calculate a reference excess chemical potential to be $\beta \mu_{\rm S}^{\rm ex}=7.3699 \pm 0.0005$ 
at $P \sigma^2/\epsilon=7.94$ and $k_B T/\epsilon=0.45$ [and where $\beta = (k_B T)^{-1}$], a $T$ which should fall between the 
two melting temperatures, using 
the Frenkel-Ladd method~\cite{ahmad2,frenkel2}.  This method requires simulations at constant $N$ and $\rho$, which we find to be
$\rho\,\sigma^2=0.907$
at this state point, with an uncertainty of $\pm$0.002.  
The ideal gas contribution to the chemical potential is $\beta \mu_{\rm id} = \ln{ \Lambda^2 \rho}$, where 
$\Lambda$ is the de Broglie wavelength.

For the liquid, we determine $\mu_{\rm L}$ at $P \sigma^2/\epsilon=7.94$ and $k_B T/\epsilon=0.7$ using two thermodynamic paths.
First, as in Ref.~\cite{ahmad2}, we integrate the equation of state along the $k_B T/\epsilon=0.7$ supercritical isotherm after 
fitting it to a phenomenological fitting model~\cite{barboy,noro}.  Second,
as a check and to have a more independent estimate of the uncertainty, we determine the enthalpy difference between
our system and the hard disk system as modeled by the equation of state~\cite{henderson,boublik},
\begin{equation}
\frac{P}{\rho k T}=\frac{1+\eta^2/8}{\left(1-\eta\right)^2},
\label{hdeos}
\end{equation}
where $\eta=\rho \pi \sigma^2/4$ is the area packing fraction.  It is somewhat straightforward to obtain 
at arbitrary state points both the hard-disk enthalpy 
$H_{\rm HD}=N P/\rho + N k_B T$
and chemical potential,
\begin{equation}
\mu_{\rm HD}(\rho) = f_{\rm id} + P/\rho +  k_B T \int_{0}^{\rho}\left( \frac{P}{\rho k_B T} -1 \right) \frac{d\rho}{\rho},
\end{equation}
where $f_{\rm id} = k_B T (\ln{ \Lambda^2 \rho -1) }$ is the ideal gas Helmholtz free energy per particle. 

\begin{figure}[ht]
\centering\includegraphics[clip=true, trim=0 0 0 0, height=6.0cm, width=8.25cm]{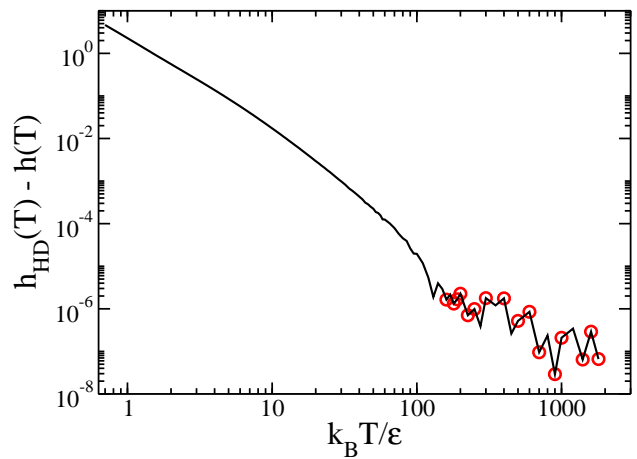}
\caption{The absolute value of the integrand of Eq.~\ref{eqintHD}, where the quantity 
$h(T)= H(T)/(Nk_BT^2) \times \epsilon/k_B$ 
at $P \sigma^2/\epsilon=7.94$, and $h_{\rm HD}(T)$ is the analogous quantity for the hard disk model.  
Circles indicate points for which the calculated enthalpy difference is negative,
and indicate the $T$ at which the integrand is essentially noise.}
\label{figintHD}
\end{figure}

The chemical potential for our system can then be written as,
\begin{equation}\label{eqintHD}
\frac{\mu_{\rm L}(T_0)}{k_B T_0}=\frac{\mu_{\rm HD}(T_0)}{k_B T_0} + \int_{T_0}^{T_{\infty}} \frac{\left(H_{\rm HD}(T)-H(T)\right)}{N k_B T^2} dT,
\end{equation}
where we have assumed that $H_{\rm HD}(T_{\infty})=H(T_{\infty})$ and used the relation,
\begin{equation}
\frac{\mu(T_{2},P)}{k_B T_{2}} = \frac{\mu(T_{1},P)}{k_B T_{1}} - \int_{T_{1}}^{T_{2}} \frac{H(P,T)}{N k_B T^2} dT.
\label{free_energy}
\end{equation}
The integrand in Eq.~\ref{eqintHD} is plotted in Fig.~\ref{figintHD} and we see that beyond $k_B T /\epsilon\approx 200$, the integrand 
is essentially noise.  We evaluate the integral both directly and with a change in variable of $\tau=\ln{T}$ using different interpolation orders
to values of $k_B T_\infty/\epsilon$ ranging from 200 to 2000.
For the hard disks at  $k_B T_0/\epsilon=0.7$ and $P_0\sigma^2/\epsilon=7.94$, 
$\beta \mu_{\rm HD}^{\rm ex}=12.855287$ (excess chemical potential).

Combining results from the two different thermodynamic routes, we obtain the excess chemical potential for our liquid
at $T_0$ and $P_0$, where the liquid density is $\rho \sigma^2=0.893\pm 0.003$, 
to be $\beta \mu_{\rm L}^{\rm ex}=13.323 \pm 0.006$.  

\begin{figure}[ht]
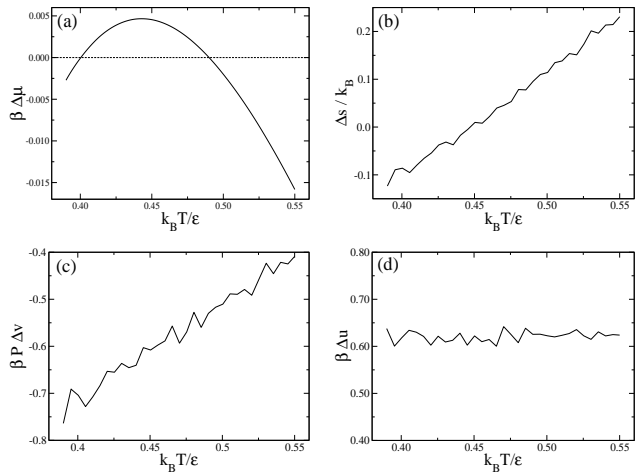

\subfigure{
\centering\includegraphics[clip=true, trim=0 0 0 0, height=3.0cm, width=4.0cm]{fig4a}
\label{chempotdiff}
}
\subfigure{
\label{entropy_liquid-crystal}
\centering\includegraphics[clip=true, trim=0 0 0 0, height=3.0cm, width=4.0cm]{fig4b}
}
\subfigure{
\label{pv_liquid-crystal}
\centering\includegraphics[clip=true, trim=0 0 0 0, height=3.0cm, width=4.0cm]{fig4c}
}
\subfigure{
\label{energy_liquid-crystal}
\centering\includegraphics[clip=true, trim=0 0 0 0, height=3.0cm, width=4.0cm]{fig4d}
}
\caption{Differences in thermodynamic quantities between the L and S 
phases as a function of $T$ at 
$P\sigma^2/\epsilon=7.94$, where $\Delta q\equiv  q_L - q_S$ for a per particle quantity $q$.
Quantities considered are (a) chemical potential $\Delta \mu$, where a negative value indicates that the 
liquid is the stable phase, (b) entropy $\Delta s$, (c) the mechanical contribution to the enthalpy 
$P\,\Delta v$ and (d) potential energy $\Delta u$. 
Below $T\approx 0.39$, the S phase no longer shows appreciable metastability. }
\end{figure}

Having obtained a value of the chemical potential at reference temperatures at $P\sigma^2/\epsilon=7.94$ for
both L and S, we use Eq.~\ref{free_energy} to determine the
difference in chemical potential, $\beta \Delta \mu \equiv \beta \mu_{\rm L}(T) - \beta \mu_{\rm S}(T)$ 
as a function of $T$, which we plot in Fig.~\ref{chempotdiff}.  The figure shows two $T$ at which crossing of zero occurs,
which is required for inverse melting to occur.  However, given the uncertainties in calculating the chemical potential and the small
value of $\beta \Delta \mu$, it is entirely possible that the liquid does not crystallize along this pressure at all.
Therefore, when we amplify the inverse melting effect below, it is necessary to check the effect by 
complementary methods. 

Fig.~\ref{entropy_liquid-crystal} shows the entropy of the crystal becoming increasingly  larger than that of the liquid for
$T$ decreasing below $k_B T/\epsilon\approx 0.45$, which is required for crystallization upon heating past the lower
of the two coexistence $T$.  Fig.~\ref{pv_liquid-crystal} shows that the volume contribution to the enthalpy of the crystal in this range
also becomes increasingly larger than the liquid's as $T$ decreases, which tends to destabilize the crystal
with respect to the liquid.   Fig.~\ref{energy_liquid-crystal} shows that the energetic driving force for phase 
transformation does not change with $T$.

Having obtained at $P_0$ two coexistence temperatures $T_{m1}=0.400345 \epsilon/k_B$ and $T_{mHigh}=0.490054 \epsilon/k_B$, we carry 
out a Gibbs-Duhem integration~\cite{kofke1,kofke2}, as in Ref.~\cite{ahmad2}, 
of the Clausius-Clapeyron equation that describes the slope in the $P$-$T$ plane
of the coexistence line,
\begin{equation}
\frac{dP}{dT} = \frac{\Delta s}{\Delta v} = \frac{\Delta h}{T \Delta v},
\label{gibbs_duhem}
\end{equation}
where $\Delta s$ is the molar entropy difference, $\Delta h$ is the molar enthalpy difference, 
and $\Delta v$ is the molar volume (area in 2D) difference between the two coexisting phases.
To test the accuracy of the integration, we carry it out twice, starting from the state point
$(P_0,T_{m1})$, labelled $C_1$ in Fig.~\ref{inverse_melting_slope-dp}, and increasing $T$ until 
$T_{mHigh}$, and again from $(P_0,T_{mHigh})$ down in temperature.  The overlapping results 
for the coexistence line are shown in Fig.~\ref{inverse_melting_slope-dp}.  The uncertainty in the position
of the line is predominantly due to the uncertainty in calculating the reference entropy of the liquid.

\subsection{Hamiltonian Gibbs-Duhem integration}

After determining the coexistence curve that exhibits inverse melting, we use Hamiltonian Gibbs-Duhem integration to find the potential parameters $(\epsilon_1, b, c)$ that increase the range of inverse melting. This technique allows one to find a coexistence point for a system governed by a potential energy $U_B$ starting from a known coexistence point for the system defined by 
potential energy $U_A$. 
The starting point is to introduce a potential that depends on a coupling parameter $\lambda$, which we 
choose to be ~\cite{vega2,romano},
\begin{equation}\label{ulambda}
U(\lambda) = \lambda U_B + (1-\lambda) U_A.
\end{equation}
As $\lambda$ changes from zero to one, the potential continuously transforms from $U_A$ to $U_B$.  In our case,
$U_A$ is determined by the SSSW pair potential using the original parameters, while $U_B$ is given 
by the SSSW potential with a different set of parameters.

Ref~\cite{singer} has shown that the generalized Clapeyron equations for two coexisting phases I and II at constant pressure and temperature can be written as, respectively,

\begin{eqnarray}
\label{hgdi_constantP}
\left. \frac{dT}{d\lambda} \right|_P &=& T \frac{\left<\partial u_{II}/\partial \lambda\right>_{NPT\lambda} - \left<\partial u_{I}/\partial \lambda\right>_{NPT\lambda}} {h_{II} - h_{I}} \\
\left. \frac{dP}{d\lambda} \right|_T &=& - \frac{\left<\partial u_{II}/\partial \lambda\right>_{NPT\lambda} - \left<\partial u_{I}/\partial \lambda\right>_{NPT\lambda}} {v_{II} - v_{I}}
\label{hgdi_constantT}
\end{eqnarray}
where $\partial u_{I}/\partial \lambda$, given in Eq.~\ref{ulambda}, is the quantity $U_B - U_A$ per particle for phase I,
$h_I$ is its per particle enthalpy and $v_I$ its per particle volume.
Similarly for phase II. $\left< . \right>_{NPT\lambda}$ indicates an average in the $NPT$ ensemble when the system is governed
by $U(\lambda)$.
In principle, by applying this technique to many coexistence points, one can obtain the phase diagram of a 
new model potential starting from a known phase diagram of another model.
 
To simplify finding the optimized parameters that can increase the inverse melting, we implement the Hamiltonian Gibbs-Duhem integration at constant temperature, given in Eq.~\ref{hgdi_constantT},  first for only two coexistence points on the inverse melting curve, labelled $C_1$ and $C_2$ in Fig.~\ref{inverse_melting_slope-dp}.
As a convenient measure of the effectiveness with which a change in the pair potential increases the region in the $P$-$T$ plane over which inverse
melting occurs, we use the slope $M = (P_{C_2} - P_{C_1})/(T_{C_2}-T_{C_1})$.  For example, if changing the potential causes $M$ to increase, then the pressure range of inverse melting increases.
We vary $\epsilon_1$, $b$ and $c$ independently to determine which parameter most effectively increases $M$. 
The two original coexistence points  that we use to study $M$ as a function of potential parameters are $C_1= \left\{ k_B T_{m1}/\epsilon=0.400345, P_0\,\sigma^2/\epsilon=7.94 \right\}$ and $C_2=\left\{k_B T_{m2}/\epsilon=0.425345, P_{m2}\,\sigma^2/\epsilon=7.98906\right\}$.
As a potential parameter is varied, the coexistence $P$ will change, causing $M$ to increase or decrease.

\subsection{Biased Monte Carlo simulations}

Once the coexistence line has been recalculated for a new set of model parameters, we choose a $P$-$T$ state point
on the coexistence line to evaluate the distribution of density fluctuations.  
This calculation is necessary to compute the free energy barrier between L and S, and to verify a coexistence point
by a different method.
The probability distribution function
for the density, determined at conditions of constant $T$ and $P$, defines
the conditional (or Landau) Gibbs free energy,
\begin{equation}
\Delta G(T,P;\rho) = G(T,P;\rho) -G_0 = -k_B T \ln[P_r(\rho)],
\end{equation}
where $P_r(\rho) \, d\rho$ is the probability of finding the system with density between $\rho$ and $\rho + d\rho$ and $G_0$ is a constant that 
ensures that the average of $G(T,P;\rho)$ gives the equilibrium Gibbs free energy $G(T,P)$.
For a finite system at a first order coexistence point, there should be two peaks of equal areas in $P_r(\rho)$.  If the shapes of the peaks
are similar, the two resulting minima in $\Delta G(T,P;\rho)$ will have the same value.  The barrier between these minima arises 
from the work required to form the transition state, which for a large enough periodic system amounts to creating two interfaces that span
the width of the simulation box.

To ensure good sampling of $\rho$, we use the umbrella sampling MC simulation~\cite{frenkel1} carried out by NPT simulations to calculate  
$G(T,P;\rho)$. 
To implement umbrella sampling, we add the following constraint potential $U_c$, 
\begin{equation}
U_c(\rho) = \frac{k}{2}(\rho-\rho_0)^2,
\label{biasing_potential}
\end{equation} 
to the system potential energy. The biasing potential will force a given simulation to sample densities in the vicinity of $\rho_0$. $k$ is a constant that controls the range of sampled densities.  We use simulation windows with equally spaced values of $\rho_0$, and perform two sets of
simulations with $N=2082$ in a rectangular (two squares) simulation box (using isotropic scaling to maintain $P$), 
one with $k=640000\epsilon/ \sigma^4 $ and again with 
$k =1280000 \epsilon/ \sigma^4$.  We convert  the probability distribution from the constrained ensemble $P_{rc}(\rho)$ to the $NPT$ ensemble via
$ P_r(\rho) \propto  \exp{\left[\beta  U_c(\rho) \right]} P_{rc}(\rho)$.  The pieces of $\Delta G(T,P;\rho)$ 
determined near each $\rho_0$ can be combined by essentially shifting each to produce a smooth $\Delta G(T,P;\rho)$ for the entire density range.
We use MBAR~\cite{shirts} to accomplish this.

There will necessarily be some error in calculating coexistence conditions at which we perform umbrella sampling.
To more precisely locate the coexistence pressure, we reweight the $\Delta G(T,P;\rho)$ curve by applying a pressure shift,
\begin{equation}
\beta G(T,P';\rho) = \beta G(T,P_0;\rho) + \frac{N\beta \Delta P}{\rho} + c, 
\label{reweight}
\end{equation}
where $c$ is a constant related to normalization. The corrected coexistence pressure is then $P^\prime = P_0 + \Delta P$, where $P_0$ 
is the original coexistence pressure at which the constrained simulations are performed and  
$\Delta P$ is the pressure shift that brings the two minima in $\Delta G(T,P_0;\rho)$ to the same level. 

To distinguish the S and L  phases in a visualization of the configurations produced, we make used of the Steinhardt bond order parameters
based on spherical harmonics~\cite{steinhardt} as was done in Ref.~\cite{ahmad2}.  

\subsection{Analysis of long range correlations}

In order to distinguish the liquid, crystal and hexatic phases in two dimensions, one typically measures or calculates 
translational and orientational correlation functions~\cite{gasser}.  For translations, in addition to the 
radial distribution function $g(r)$, we calculate,
\begin{equation}
G_{\vec{g}}(r)= \left< \exp{\left(i \vec{g} \cdot \vec{r}_j \right)} \right>,
\label{trans-corr}
\end{equation}
where we average the result over reciprocal lattice vectors $\vec{g} = \hat{x} \, 2\pi /a$
and $\vec{g} = \hat{y} \, 2\pi /a$, 
$a$ is the expected lattice spacing in the S phase for the density studied, 
$\vec{r}_j$ with magnitude $r$ is the position of particle $j$ relative to an origin taken to be one of the particle positions,
and $\left< . \right>$ indicates an ensemble average over origins and particles $j$. 
For orientational order, we use,
\begin{equation}
G_4(r)= \left< q_4(\vec{r})\,q_4^*(\vec{0}) \right>,
\end{equation}
\begin{equation}
q_4(\vec{r}_j) = \frac{1}{N_j}\sum_{k=1}^{N_j} \exp{\left(4 i \theta_{jk} \right)},
\end{equation}
where $q_4^*$ is the complex conjugate of $q_4$,
$\theta_{jk}$ is the angle made by the bond with respect to an arbitrary but fixed axis between particle $j$ and neighbor $k$,
neighbours being those particles that are closer together than a distance of $1.24\sigma$,
and the sum is over the $N_j$ neighbors of particle $j$.

The expectation based on the KTHNY theory of melting~\cite{gasser,bernard,saija} 
in two dimensions for these functions is that  
both $G_4(r)$ and $G_{\vec{g}}(r)$ decay exponentially in the liquid phase,  
that $G_4(r)$ decays as a power law with a small exponent ($\leq 1/4$) and $G_{\vec{g}}(r)$ decays exponentially in the hexatic phase, and
that  $G_4(r)$ tends to a constant and $G_{\vec{g}}(r)$ decays slowly as a power law with a small exponent ($\leq 1/3$) in the crystal.

To detect the presence of a quasicrystal phase, we calculate the structure factor,
\begin{equation}
S(\vec{q}\,) = \frac{1}{N} \left< \rho_{\vec{q}} \,\, \rho_{\vec{q}}^\ast \right>,
\end{equation}
where,
\begin{equation}
\rho_{\vec{q}}=\sum_{i=1}^N \exp{(-i \, \vec{q} \cdot \vec{r}_i) },
\end{equation}
and $\left< . \right>$ indicates an ensemble average and $\rho^\ast$ is the complex conjugate of $\rho$.
In our periodic system, the allowed reciprocal vectors are $\vec{q} = 2\pi (n_x, n_y)/L$, where $L$ is the length of the simulation box and $n_{x,y}$
are integers.


\section{Results}

\subsection{Expanding the range in inverse melting}

\begin{figure}[h]
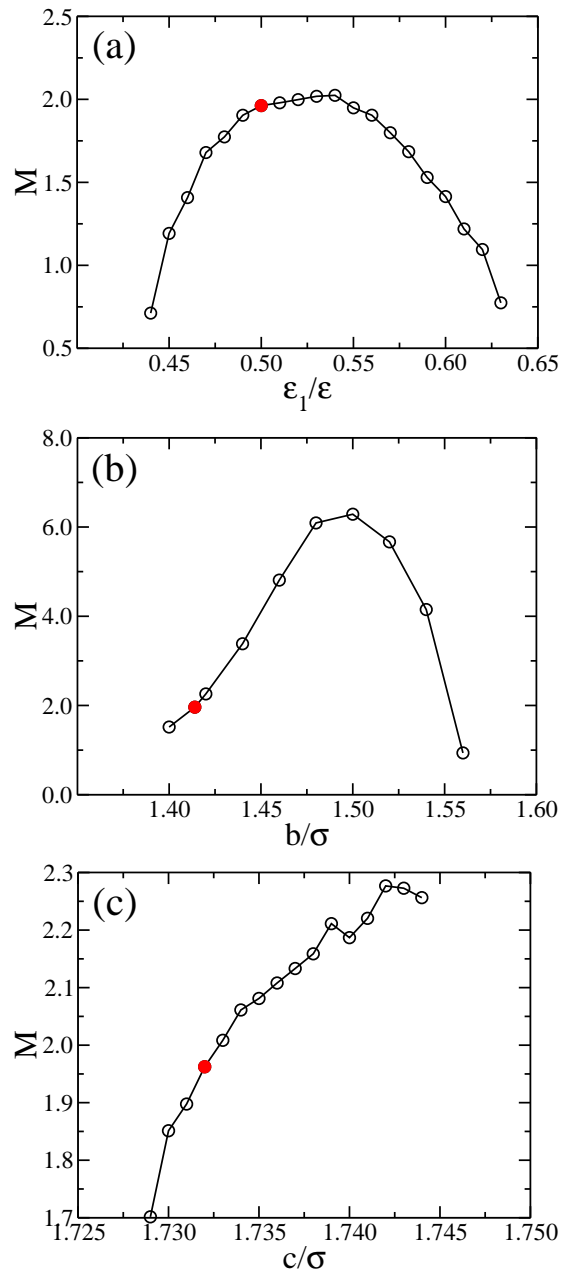

\subfigure{
\centering\includegraphics[clip=true, trim=0 0 0 0,  width=7.25cm]{fig5a}
\label{fig_delta_e}
}
\subfigure{
\label{fig_delta_b}
\centering\includegraphics[clip=true, trim=0 0 0 0,  width=7.25cm]{fig5b}
}
\subfigure{
\label{fig_delta_c}
\centering\includegraphics[clip=true, trim=0 0 0 0,  width=7.25cm]{fig5c}
}
\caption{The slope $M$ between two S-L coexistence points, 
labelled $C_1$ and $C_2$ for the original parameters in Fig.~\ref{inverse_melting_slope-dp}, changes as a function of
(a) $\epsilon_1$, (b)  b and (c) c.
A larger value of $M$ implies inverse melting occurring over a larger range of $P$.  Filled (red) circles indicate
$M$ for the original parameter values.
}
\label{figM}
\end{figure}

Each panel in Fig.~\ref{figM} shows how the 
slope $M$ between two selected points on the original S-L coexistence curve, 
$C_1$ and $C_2$ [Fig.~\ref{inverse_melting_slope-dp}]
changes when each of $\epsilon_1$, $b$ and $c$ is varied with
the other two parameters  held fixed.
A larger value of $M$ compared with the original parameters indicates an expanded range of pressures over which inverse melting should be observed.
The filled red circle in each panel represents the value of $M$ when using the original potential parameters: $\epsilon_1=\epsilon/2$, $b=\sqrt{2}\sigma$ and $c=\sqrt{3}\sigma$. From Fig.~\ref{fig_delta_e}, we conclude that $M$ is already near the maximum for the original value of $\epsilon_1$, and therefore changing this parameter will not help increase the range of inverse melting. 
On the other hand, Fig.~\ref{fig_delta_b} shows that $M$ increases by
a factor of three when $b$ is increased, greatly expanding the range of inverse melting. 
Increasing the parameter $c$ beyond the values shown in Fig.~\ref{fig_delta_c} results in losing the L phase in favor of
HDT.  Thus it appears that in this case, L-S inverse melting becomes metastable with resect to HDT.

\begin{figure}[ht]
\centering\includegraphics[clip=true, trim=0 0 0 -30, width=8.8cm]{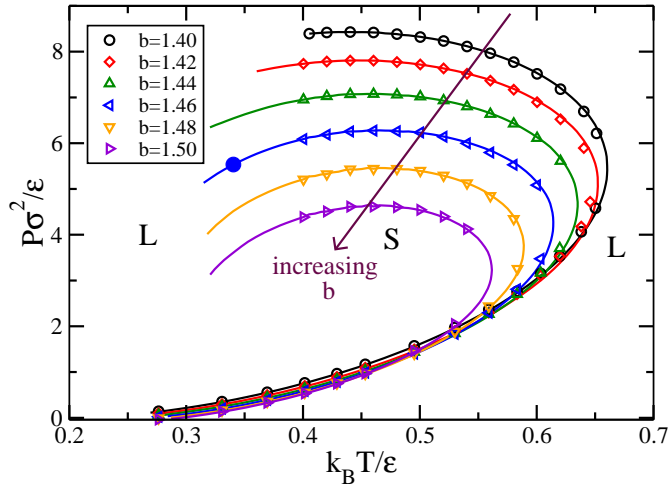}
\caption{The S-L coexistence curve for $b/\sigma$ ranges from $1.40$ to $1.50$, calculated by Gibbs-Duhem integration method (solid lines) and Hamiltonian Gibbs-Duhem integration method (discrete points).  Large filled circle indicates coexistence point at which biased Monte Carlo
simulations explicitly show a free energy barrier between the S and L phases.}
\label{inverse_melting_curves}
\end{figure}

Given that $b$ alone is the important parameter in increasing the range of inverse melting, we proceed
with a more detailed look at how the S-L coexistence curve changes with $b$.
To begin, we perform a Gibbs-Duhem integration for the original interaction parameters starting from the coexistence point
($k_B T/\epsilon=0.490345$, $P\sigma^2/\epsilon=7.94$) to obtain the full curve.
For roughly 20 points on this curve, we carry out Hamiltonian Gibbs-Duhem integration for $b/\sigma=1.40, 1.42, 1.44, 1.46, 1.48$ and $1.50$.
The results are represented by open symbols in Fig.~\ref{inverse_melting_curves}.
To check the accuracy of determining these points, we perform Gibbs-Duhem integration for each value of $b$, 
starting from $T_{mHigh}$,
as represented by the solid lines in Fig.~\ref{inverse_melting_curves}. The results obtained by the two integration methods shows a high degree of agreement. From Fig.~\ref{inverse_melting_curves}, it becomes obvious that as the potential parameter $b$ increase, the range of pressure of the inverse melting increases.  Concurrent with this change is the reduction of the S stability field.

For the analysis that follows, we focus on the SSSW potential for which $b/\sigma=1.46$ while $\epsilon_1$ and $c$ are kept at their original values.
Already at this value of $b$, the range in $P$ over which the coexistence line exhibits inverse melting is considerable.  This allows more direct methods
to confirm the phenomenon.

\subsection{Interfacial tension between S and L}

\begin{figure}[ht]
\centering\includegraphics[clip=true, trim=0 0 0 0, height=6.0cm, width=8.3cm]{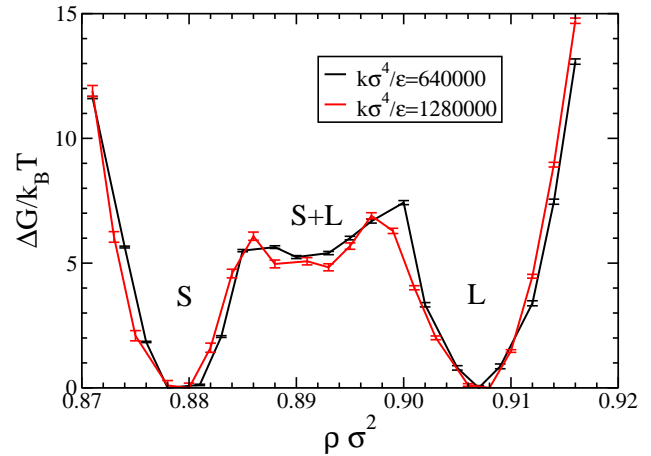}
\caption{Conditional Gibbs free energy as a function of $\rho$ calculated at temperature $k_B T/\epsilon = 0.340345$ and reweighted by Eq.~\ref{reweight}. The black curve is calculated for $k \sigma^4/\epsilon=640000$ and red for $1280000$.
}
\label{free_energy_N2082}
\end{figure}

\begin{figure}[ht]
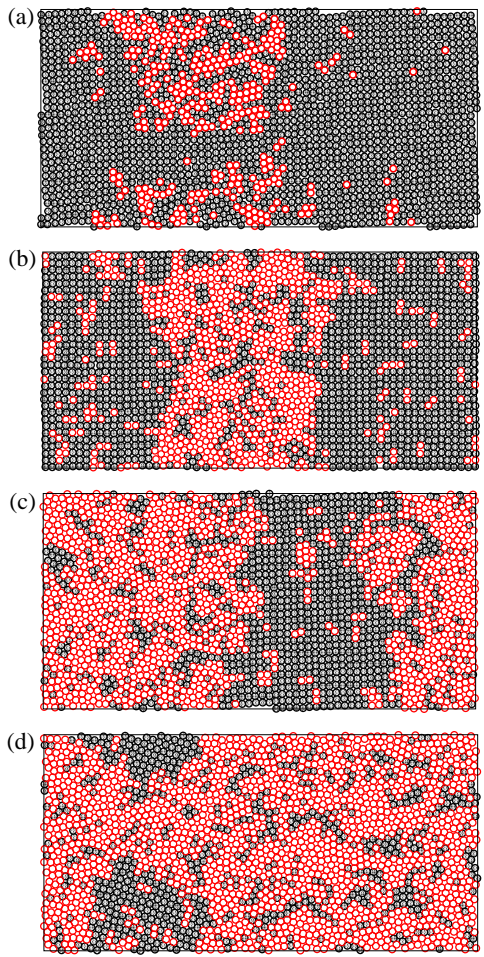

\subfigure{
\centering\includegraphics[clip=true, trim=0 0 0 0, height=3cm]{fig8a}
\label{liquid_bubble}
}
\subfigure{
\label{liquid_strip}
\centering\includegraphics[clip=true, trim=0 0 0 0, height=3cm]{fig8b}
}
\subfigure{
\label{square_strip}
\centering\includegraphics[clip=true, trim=0 0 0 0, height=3cm]{fig8c}
}
\subfigure{
\label{square_bubble}
\centering\includegraphics[clip=true, trim=0 0 0 0, height=3cm]{fig8d}
}
\caption{Snapshot of configurations taken from different windows of umbrella sampling for $N=2082$ particles. Fig.~\ref{liquid_bubble} shows a liquid bubble at $\rho_0\sigma^2=0.8862$, Fig.~\ref{liquid_strip} shows a liquid strip at $\rho_0\sigma^2=0.8887$, Fig.~\ref{square_strip} shows a square-crystal strip  at $\rho_0\sigma^2=0.8975$ and Fig.~\ref{square_bubble} shows a square-crystal bubble at $\rho_0\sigma^2=0.9000$.}
\label{bubbles}
\end{figure}

To confirm inverse melting, we report the $G(T,P;\rho)$ from a histogram of the densities sampled by a series of
biased NPT simulations with 2082 particles at the coexistence point ($k_B T/\epsilon=0.340345$, $P \sigma^2/\epsilon=5.5331$) 
for the SSSW
model for which $b/\sigma=1.46$.  This coexistence point is indicated by the large filled circle in Fig.~\ref{inverse_melting_curves}.

The results are shown in Fig.~\ref{free_energy_N2082}, where we use Eq.~\ref{reweight} to bring the free energy minima to the same level.
The pressure shifts required in this reweighting are small, $\Delta P \sigma^2/\epsilon = -0.0261$ 
for the simulations with $k \sigma^4/\epsilon=6.4 \times 10^5$
and $\Delta P \sigma^2/\epsilon = -0.0295$ 
for the simulations with $k \sigma^4/\epsilon=12.8 \times 10^5$, indicating that the errors built up during
the several step in determining the coexistence line is indeed small.  The curves show a barrier of approximately 5~$k_B T$ separating
the lower density S phase from the higher density liquid.

The shape of the barrier, generally flat with overshoots at either end, is consistent with the morphology of the separated phases.
Despite the rather diffuse interface between S and L, as noted in Ref.~\cite{ahmad2}, the system is large enough to accommodate
an isolated liquid droplet within the S phase.  This we show in Fig.~\ref{liquid_bubble}, which shows a snapshot from the biased
$NPT$ simulation with $\rho_0 \sigma^2= 0.8862$, i.e., near the overshoot occurring as the density of the system is constrained to the high
density side of the S basin in $G(T,P;\rho)$.  At higher density, Fig.~\ref{liquid_strip}, the L phase spans the width of the periodic simulation
cell.  If the width of the strip is sufficiently wide to accommodate two well formed S-L interfaces, 
then increasing the density further will not change the free energy,
as both phases are at the same chemical potential.  In our case, the system may not be large enough to accomplish this, 
as we can only claim a broad minimum near $\rho \sigma^2= 0.89$ and not a truly a flat region in the barrier,
and so we can only estimate an upper bound on the interfacial tension.  Taking the minimum of the barrier to be
$\beta \Delta G(T,P;\rho^*)=5.1\pm0.2$, we determine the interfacial tension
to be $\gamma \sigma/\epsilon= \Delta G(T,P;\rho^*)/(2 w) = 0.025 \pm 0.001$,
where $w/\sigma = 34.160$ 
is the box width and $\beta/\epsilon=0.340345^{-1}$, or $\beta \gamma \sigma = 0.075 \pm 0.003$.
Increasing the density further results first in a strip of the S phase within the liquid [Fig.~\ref{square_strip}]
and then a bubble of S [Fig.~\ref{square_bubble}] before reaching the homogeneous liquid.

\subsection{Direct simulation of freezing and melting}

\begin{figure}[ht]
\centering\includegraphics[clip=true, trim=0 0 0 0, height=6.0cm, width=8.5cm]{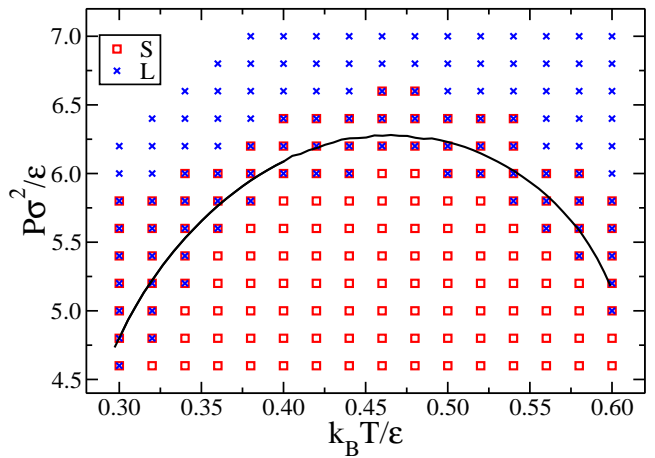}
\caption{The solid line is the S-L coexistence curve for the potential parameter $(\epsilon_1=0.5,\ b=1.46\sigma\ \text{and}\ c=\sqrt{3}\sigma$) and the grid of points is obtained from two sets of NPT MC simulations, one beginning from a perfect S configuration and the other from a liquid.
A blue x symbol represents a simulation that ended in the L phase, while a 
red open square represents a simulation that ended in the S phase.
A state point with both symbols indicates that each simulation retained its starting phase.  
}
\label{liquid_state_points}
\end{figure}

As a rough check on the portion of the S-L coexistence curve that exhibits inverse melting and to determine the extent of metastability, we preform a set of NPT simulations for the potential parameters $\epsilon_1=0.5$, $b=1.46\sigma$ and $c=\sqrt{3}\sigma$ to map out the range of metastability of L and S. For both phases, we use 1024 particles in a square box, scaling the box size isotropically to maintain $P$. We initialize the L simulations with a liquid configuration and the S simulations with a square crystal, and we run each state point for $4\times10^8$ MC steps per particle. We indicate with a blue x sign in Fig.~\ref{liquid_state_points} the state points for which simulations either retain the L phase or melt to the L phase, and with a red open square symbol the state points for which simulations either retain the S phase or crystallize to S.

From Fig.~\ref{liquid_state_points}, we see that the L phase is obtained well above the coexistence curve and the S crystal is obtained  
for state points well within 
its predicted stability field. At and near the coexistence curve, we see  both phases at every state point, indicating the stability or metastability of the two phases. We also see the tendency for points exhibiting either liquid or S metastability to track the curvature of the S-L melting line. 
For this system size, inverse melting is directly confirmed at $P\sigma^2/\epsilon = 6.0$: at low $T$, only the liquid survives; at 
$k_B T/\epsilon\approx0.46$ only S survives; and by $k_B T/\epsilon\approx0.60$, only the liquid is stable.

\begin{figure}[ht]
\centering\includegraphics[clip=true, trim=0 0 0 0, height=6.0cm, width=8.5cm]{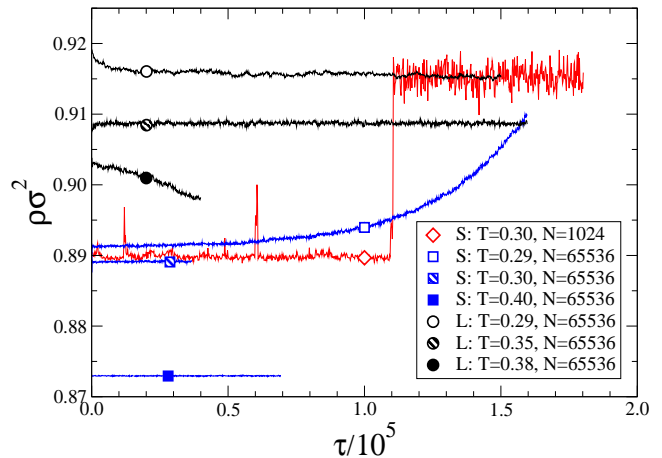}
\caption{
Density as a function of time from EDMD simulations at $P\sigma^2/\epsilon=5.6$.  Here reduced time $\tau= t\sqrt{\epsilon/(\sigma^2m)}$.  Legend gives
initial phase, $T$ and system size.  For the $N=1024$ simulation, time is reduced by a factor of 50 for ease of comparison
(i.e. its simulation time is roughly 50 times larger than the longest simulation for $N=65536$).
}
\label{edmd}
\end{figure}

To independently confirm these findings, we carry out EDMD simulations of $1024$ and $65536$ particles along the 
$P\sigma^2/\epsilon = 5.6$ isobar.  In Fig.~\ref{edmd} we plot the density as a function of time for a few $T$,
chosen to illustrate the behavior of both phases when they are stable, metastable and undergoing a phase
transformation.  The most dramatic and direct illustration of melting of S at low $T$ is for the
$N=1024$ simulation at $k_B T/\epsilon=0.30$, which started from a perfect S crystal, where 
the density exhibits a sudden increase as the system transforms from S to L.  Such a jump is
typical of first order transitions when the crystallization of the system is dominated by nucleation.
Note that time for this smaller system is reduced by a factor of 50 for plotting purposes in order to compare with
the time scales of the $N=65536$ simulations.

\begin{figure}[ht]
\subfigure{
\centering\includegraphics[clip=true, trim=0 0 0 0, height=4.0cm, width=4.0cm]{fig11a}
\label{Config_T030_SquareCryst_LiqParticles}
}
\subfigure{
\label{Config_T029_SquareCryst_LiqParticles}
\centering\includegraphics[clip=true, trim=0 0 0 0, height=4.0cm, width=4.0cm]{fig11b}
}
\subfigure{
\label{Config_T035_Liquid_SqrParticles}
\centering\includegraphics[clip=true, trim=0 0 0 0, height=4.0cm, width=4.0cm]{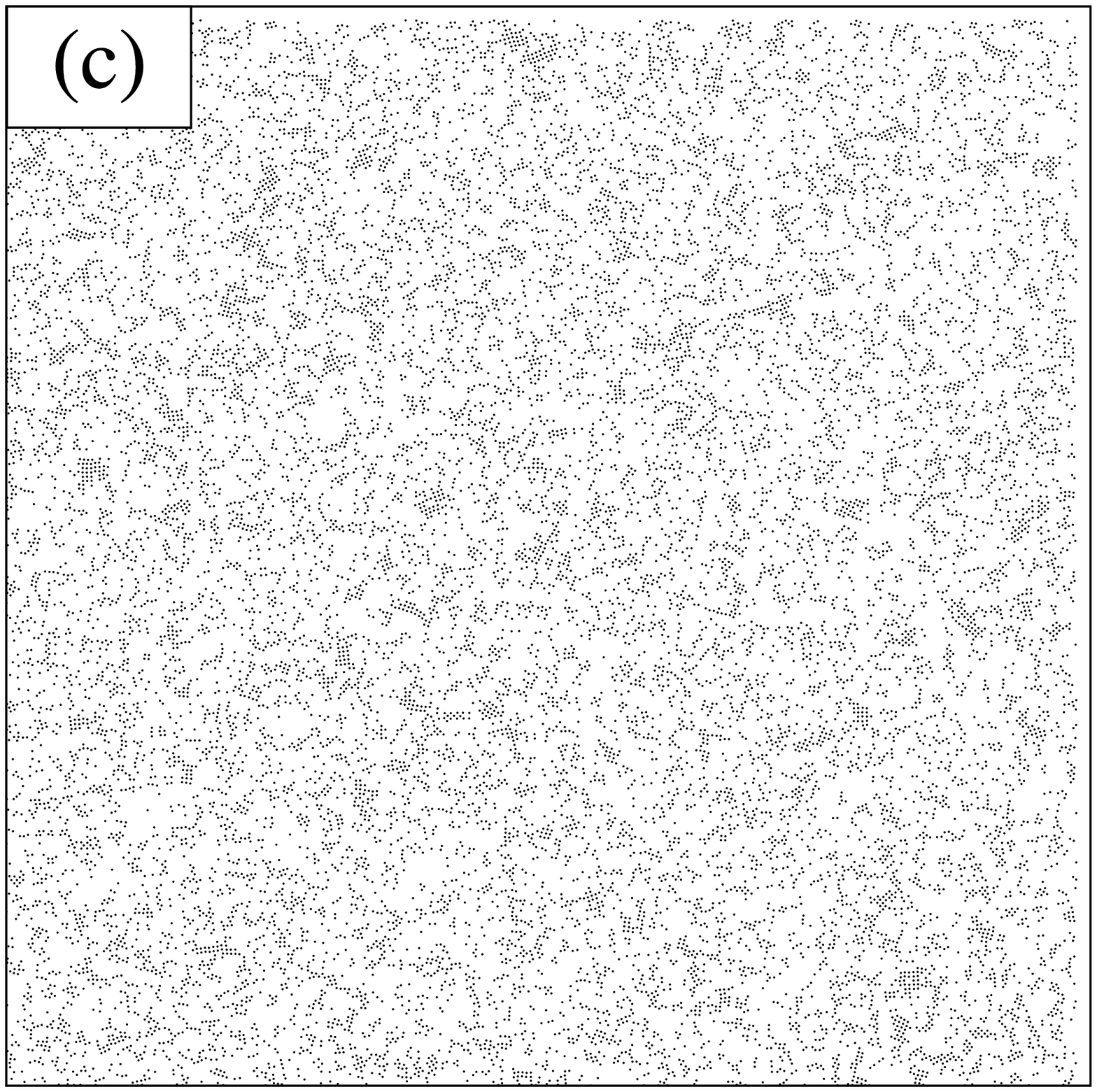}
}
\subfigure{
\label{Config_T038_Liquid_SqrParticles}
\centering\includegraphics[clip=true, trim=0 0 0 0, height=4.0cm, width=4.0cm]{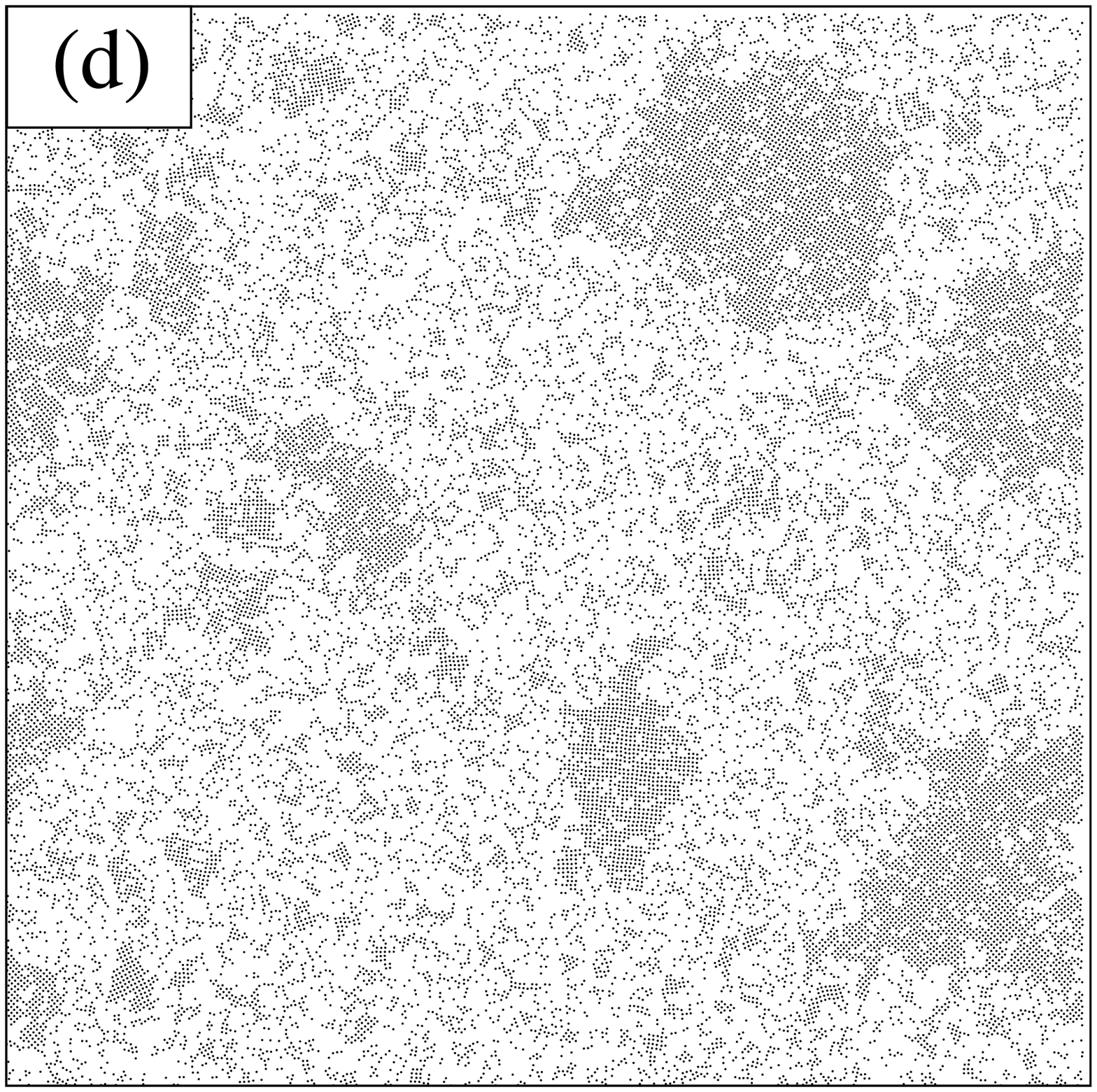}
}
\caption{
Snapshot configurations from EDMD simulations showing liquid-like particles for 
(a) the metastable S phase at $k_B T/\epsilon=0.30$ at $\tau=36741$, and
(b) the S phase melting at $k_B T/\epsilon=0.29$ at $\tau=74732$ and showing several distinct liquid domains;
and configurations showing S-like particles for
(c) the metastable L phase at $k_B T/\epsilon=0.35$ at $\tau=46459$ and 
(d) crystallizing L at  $k_B T/\epsilon=0.38$ at $\tau=40000$.
Density time series for these state points are shown in Fig.~\ref{edmd}.
}
\label{snapshots}
\end{figure}

To observe melting of S for the larger system on a reasonable time scale, we reduce the temperature to $k_B T/\epsilon=0.29$.  Here,
the slow, rather continuous increase in $\rho$ arises from crystallization being dominated by 
growth.  In Fig.~\ref{Config_T030_SquareCryst_LiqParticles} we plot dots representing the 
rather uniformly distributed locations of L-like particles within the metastable S phase for a snapshot configuration 
at $k_B T/\epsilon=0.30$
and reduced time $\tau= t\sqrt{\epsilon/(\sigma^2m)}= 36741$.
Fig.~\ref{Config_T029_SquareCryst_LiqParticles} shows distinct domains of the L phase appearing as 
S melts at $k_B T/\epsilon=0.29$ (snapshot taken at $\tau=74732$), which is consistent with the first order nature of the
transition.
The time series for S at $k_B T/\epsilon=0.40$ is representative of the stable S phase.

Addressing the liquid, we show in Fig.~\ref{edmd} density time series for three state points: 
$k_B T/\epsilon=0.29$, where L is thermodynamically stable; 
$k_B T/\epsilon=0.38$, where L is
unstable and the time series decays to lower $\rho$ as the S phase forms; and 
$k_B T/\epsilon=0.35$, where the time series is stable and, according to our calculated phase boundaries, L is metastable. 
Snapshots from the $k_B T/\epsilon=0.35$ ($\tau=46459$) and $k_B T/\epsilon=0.38$ ($\tau=40000$) 
simulations showing only S-like particles are plotted in Figs.~\ref{Config_T035_Liquid_SqrParticles}
and \ref{Config_T038_Liquid_SqrParticles}, respectively.  Similarly to the case of crystal melting,
we see distinct domains of the stable phase surrounded by the metastable phase in Fig.~\ref{Config_T038_Liquid_SqrParticles}.

Encouraged by these EDMD results, we perform additional EDMD simulations of $N=65536$ particles 
for the model with $b=1.46\sigma$ (which exhibits strong inverse melting)
and the original model with $b=\sqrt{2}\sigma$ (where inverse melting is at best very weak), and report the following.
For $b=1.46\sigma$, and $P \sigma^2/\epsilon=5.6$ starting from the L phase,  simulations for $k_B T/\epsilon \ge 0.56$ remain as L,
for  $0.40 \le k_B T/\epsilon \le 0.55$ transform to S, and for $k_B T/\epsilon \le 0.38$ remain as L.
Again for $b=1.46\sigma$, and $P \sigma^2/\epsilon=5.6$ but starting from the S phase, simulations for $k_B T/\epsilon \ge 0.58$ transform to L,
for $0.36 \le k_B T/\epsilon \le 0.56$ remain as S, and for $k_B T/\epsilon \le 0.34$ transform to L.  We also note that for $k_B T/\epsilon \le 0.31$
the energy of the liquid is lower than that of S.  These results are consistent with the phase diagram calculations and also point to the role 
of a lower potential energy of L with respect to S as a contributing factor in enhancing inverse melting in the $b=1.46\sigma$ model.

For the original $b=\sqrt{2}\sigma$ model (again with $N=65536$), 
it is more difficult for direct EDMD simulations to confirm inverse melting and we thus start
simulations with a system that is half S and half L to make confirmation possible.
At $P \sigma^2/\epsilon=7.7$, the system transforms to HDT for $k_B T/\epsilon \le 0.37$, appears to contain S, HDT and L 
at $k_B T/\epsilon = 0.38$ (which is close to the triple point), converts to S for $0.39 \le k_B T/\epsilon \le 0.51$ and converts to L
for $k_B T/\epsilon \ge 0.52$.
At $P \sigma^2/\epsilon=7.8$ inverse melting is confirmed. The system transforms to HDT for $k_B T/\epsilon \le 0.38$,  
converts to L for $0.39 \le k_B T/\epsilon \le 0.41$,
converts to S for $0.42 \le k_B T/\epsilon \le 0.48$
and converts to L for $k_B T/\epsilon \ge 0.50$.  We note that at $k_B T/\epsilon = 0.39$, interestingly, we observe the appearance of the 
HDT phase prior to full melting.
At $P \sigma^2/\epsilon=7.9$, the S phase is lost.  The system transforms to HDT for $k_B T/\epsilon \le 0.38$ and to L for
$k_B T/\epsilon \ge 0.39$.  Similarly, at $P \sigma^2/\epsilon=8.0$, the system transforms to HDT for $k_B T/\epsilon \le 0.39$ and to L for
$k_B T/\epsilon \ge 0.40$.  For all these state  points for the original $b=\sqrt{2}\sigma$, the potential energy of the liquid is higher than that of S.

\subsection{Ruling out hexatic and quasicrystal phases}

In two dimensions, a crystal possesses medium-range translational order and long-range orientational order.
Furthermore, there is the possibility that a crystal melts in two dimensions via a hexatic phase, which retains
medium-range orientational order before encountering the liquid, at which point orientational order is only short-range.
Additionally, in Figs.~\ref{bubbles} and \ref{snapshots} we see that, based on our bond-order parameter criteria
for identifying crystal-like and liquid-like particles, there are a large number of defects within each phase, i.e.,
many S-like particles in the L phase and vice versa.  

\begin{figure}
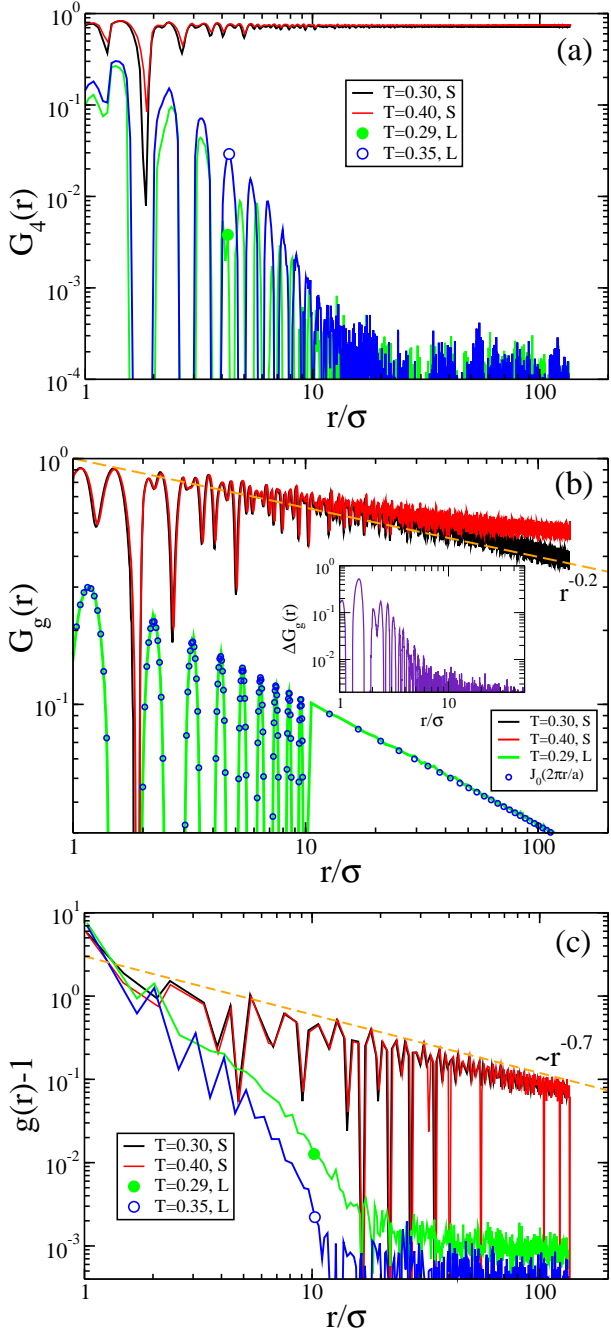

\subfigure{
\centering\includegraphics[clip=true, trim=0 0 0 0, width=8.cm]{fig12a}
\label{G4_N65536}
}
\subfigure{
\label{Gg_N65536}
\centering\includegraphics[clip=true, trim=0 0 0 0, width=8.cm]{fig12b}
}
\subfigure{
\label{gr_N65536}
\centering\includegraphics[clip=true, trim=0 0 0 0, width=8.cm]{fig12c}
}
\caption{Orientational (a) and translational (b) correlation functions as well as (c) the peaks of $g( r)-1$ for 
the $N=65536$ system at 
$P\sigma^2/\epsilon=5.6$ for the 
S phase at $k_B T/\epsilon=0.30$, 
S at $k_B T/\epsilon=0.40$, 
L at $k_B T/\epsilon=0.29$ and 
L at $k_B T/\epsilon=0.35$.  
In panel (a) $G_4(r )$ reaches a constant for S, while decaying exponentially for L.
In panel (b) $G_{\vec{g}}(r )$ 
decays as a power law for S.  For L, $G_{\vec{g}}(r )$ is described by a Bessel function $J_0(2 \pi r/a)$, which is the 
analytic result for a random system.  For L and $J_0(2 \pi r/a)$, we plot only the peaks for $r> 10\sigma$.
The inset shows exponential decay in $G_{\vec{g}}(r )$ for L 
once the vector $\vec{g}$ is chosen to align with local environments and after $J_0(2 \pi r/a)$ is subtracted.
In (c), $g( r)$ decays exponentially in the liquid, and as a power law with exponent $\sim 0.7$ for the S phase.
}
\label{g4gg}
\end{figure}

To clarify the range of order, we focus on two state points for each phase near the low $T$ melting point
along $P\sigma^2/\epsilon=5.6$: one for which the phase is thermodynamically stable and the other for which
it is metastable, given our calculated phase boundaries.  We choose
S at $k_B T/\epsilon=0.30$ (metastable), 
S at $k_B T/\epsilon=0.40$ (stable), 
L at $k_B T/\epsilon=0.29$ (stable) and 
L at $k_B T/\epsilon=0.35$ (metastable), all for the $N=65536$ for which the time series are plotted in Fig.~\ref{edmd}.
It is true that we have not quantified the effect of system size on the location of the phase boundaries, but the EDMD
simulations themselves confirm that what we deem as metastable is not far from being unstable.  
For each state point we calculate $G_4( r )$, $G_{\vec{g}}( r)$, $g( r)$ and also $S(\vec{q})$, which is calculated from a single configuration taken from the time series.
In Fig.~\ref{g4gg}(a), we plot the orientational correlation function $G_4( r )$ and see the expected behavior:
the L phase decorrelates within ten particle diameters while S remains correlated at long range as 
$G_4( r)$ approaches a constant close to unity.  Neither S nor L exhibit behavior in $G_4( r )$ that can
be interpreted as hexatic-like.

From the translational correlation functions plotted in Fig.~\ref{g4gg}(b) we see that for the 
S phase, $G_{\vec{g}}( r)$ decays as a power law with an exponent of roughly $\sim 0.1 - 0.2$, which is smaller  in magnitude 
than $1/3$, the value expected for triangular 2D crystals near the transition to the hexatic phase.
For the liquid, $G_{\vec{g}}( r)$ is smaller in magnitude than for S and oscillates about zero, 
and the peaks decay as a power law with an exponent equal to $1/2$.  
While this power-law decay is perhaps at first surprising, and indeed the same behavior 
has been observed in experiments on colloids~\cite{brodin}, it is not an indication of quasi-long-range order.
Rather, if one calculates $G_{\vec g}( r)$ by averaging over uniformly distributed orientational environments, then one obtains,
\begin{equation}
G_{\vec g}(r )= \frac{1}{\pi} \int_{0}^{\pi} \cos\bigg(\frac{2\pi r}{a} \cos(\phi)\bigg) d\phi = J_{0}\bigg(\frac{2\pi r}{a}\bigg),
\end{equation}
where $\phi$ is the angle between $\vec{g}$ and $\vec{r}_j$ in Eq.~\ref{trans-corr}, and $J_0( r)$ is the Bessel function of the first kind. 
The blue open circles in Fig.~\ref{Gg_N65536} represent $J_{0}(2\pi r/a)$ 
averaged over the $a$ values of the liquid configurations at $T=0.29 k_B/\epsilon$ used to
calculate $G_{\vec g}(r )$. 
The result shows a complete agreement with the $G_{\vec g}( r)$ of the liquid phase.
We note that for the liquid curves in Fig.~\ref{g4gg}(b), for clarity, we only plot for $r > 10\sigma$
points corresponding to local peaks in $G_{\vec{g}}( r)$ and $J_0( r)$.
$J_0( r)$ decays as $1/\sqrt{r}$, which accounts for the observed power law.  Subtracting $J_{0}(2\pi r/a)$ from
$G_{\vec g}( r)$ gives essentially noise and a correlation length of zero.

In calculating $G_{\vec{g}}(r )$ so far, $\vec{g}$ is constant, i.e., the reference system is that of the simulation box.  This makes sense for 
a crystal, but choosing a lattice vector for the liquid must take into account local ordering.
We therefore employ the method whereby every time we select a particle to be an 
origin, we use each of its closest four neighbors in turn to define the $x$ direction, and then average over the four $\vec{g}=2\pi/a \, \hat{x}$ 
reciprocal lattice vectors for that origin.  Doing so catches local translational ordering in the absence of a global orientation.  
The result is a larger correlation at small $r$ for L, but nonetheless $G_{\vec{g}}(r )$ rapidly approaches the Bessel function result.  
To more clearly see the decay in correlation, we plot in the inset of Fig.~\ref{Gg_N65536} the quantity 
$\Delta G_{\vec{g}}(r ) \equiv G_{\vec{g}}(r ) - J_{0}(2\pi r/a)$, where $G_{\vec{g}}(r )$ now takes into account local orientation.
The exponential decay in this case has a somewhat smaller length scale than what is seen in the orientational correlations for L,
but at least the exponential decay is observed.  

By contrast, a plot of the peaks of $g(r )$ in Fig.~\ref{g4gg}(c) distinguishes in a more straightforward way  between the liquid and crystal
in terms of the range of order.
For the liquid, $g(r )$
shows an exponential decay  with
a similar length scale present in $G_4( r )$.
For S, there is a power-law decay, with an exponent of approximately 0.7, significantly larger than the exponent for $G_{\vec{g}}( r)$.

\begin{figure}[!ht]
\subfigure{
\label{sqnum4}
\centering\includegraphics[clip=true, trim=0 0 0 0, height=4.0cm, width=4.0cm]{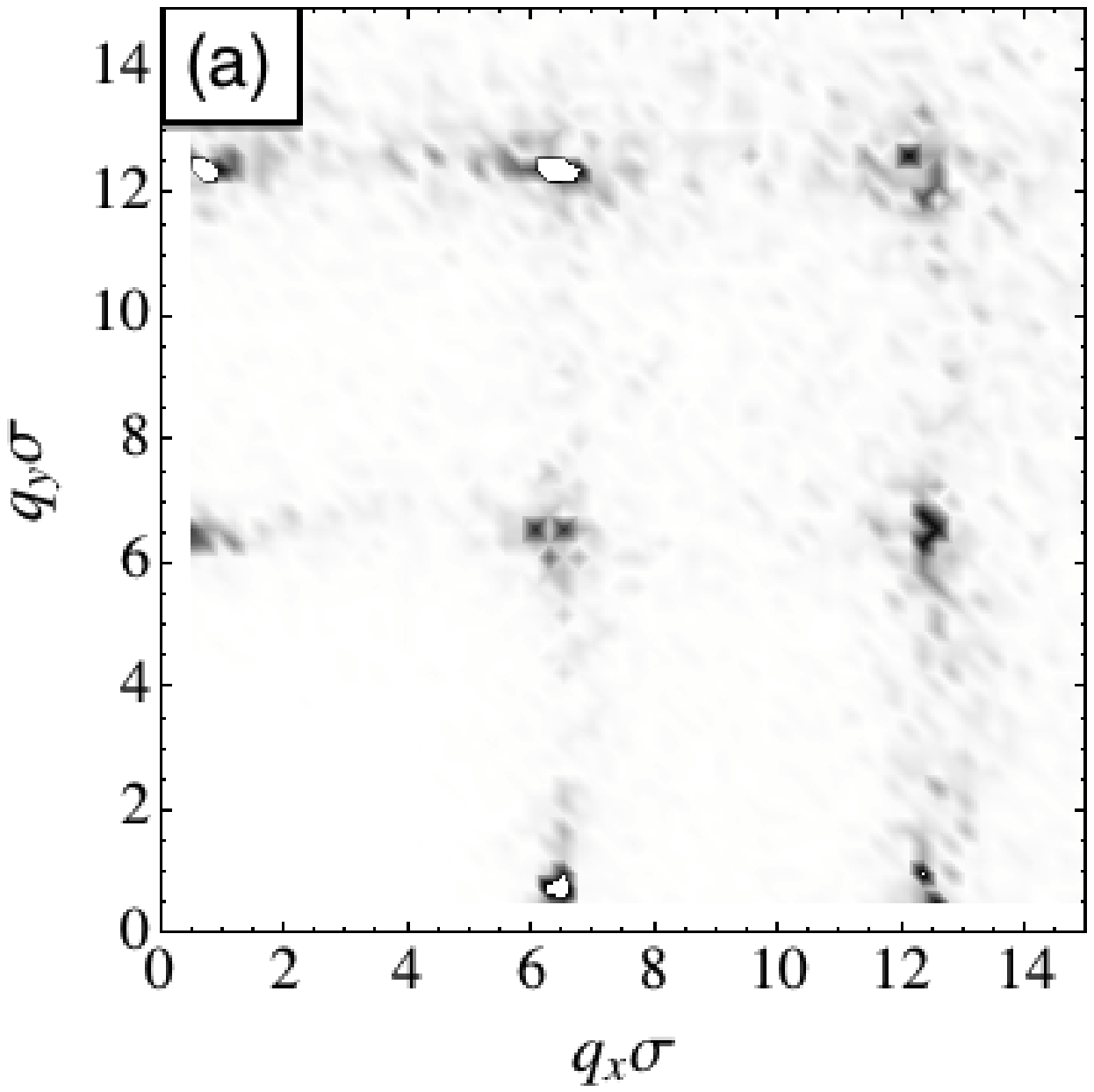}
}
\subfigure{
\label{sqnum7}
\centering\includegraphics[clip=true, trim=0 0 0 0, height=4.0cm, width=4.0cm]{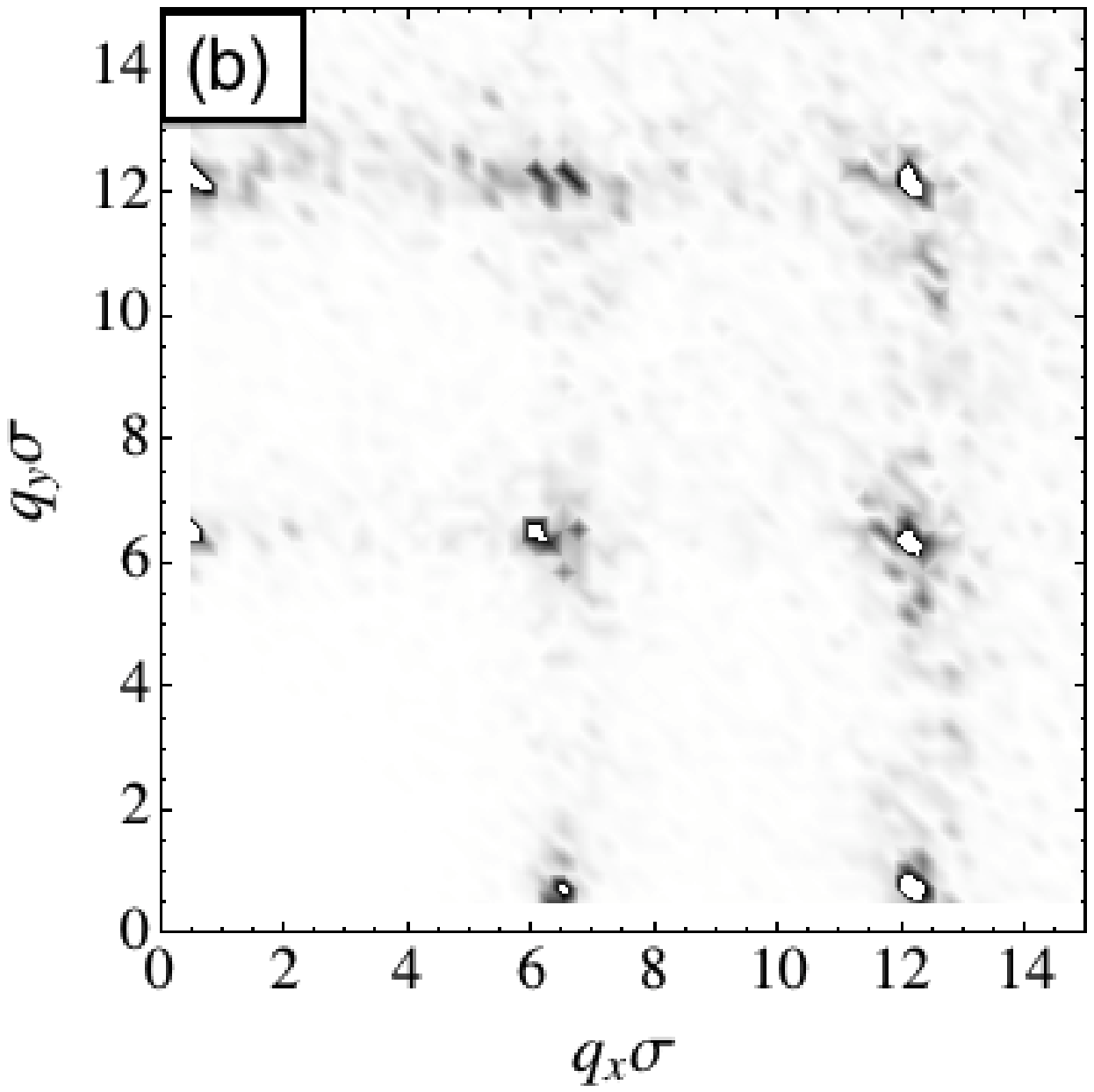}
}
\subfigure{
\label{sqnum2}
\centering\includegraphics[clip=true, trim=0 0 0 0, height=4.0cm, width=4.0cm]{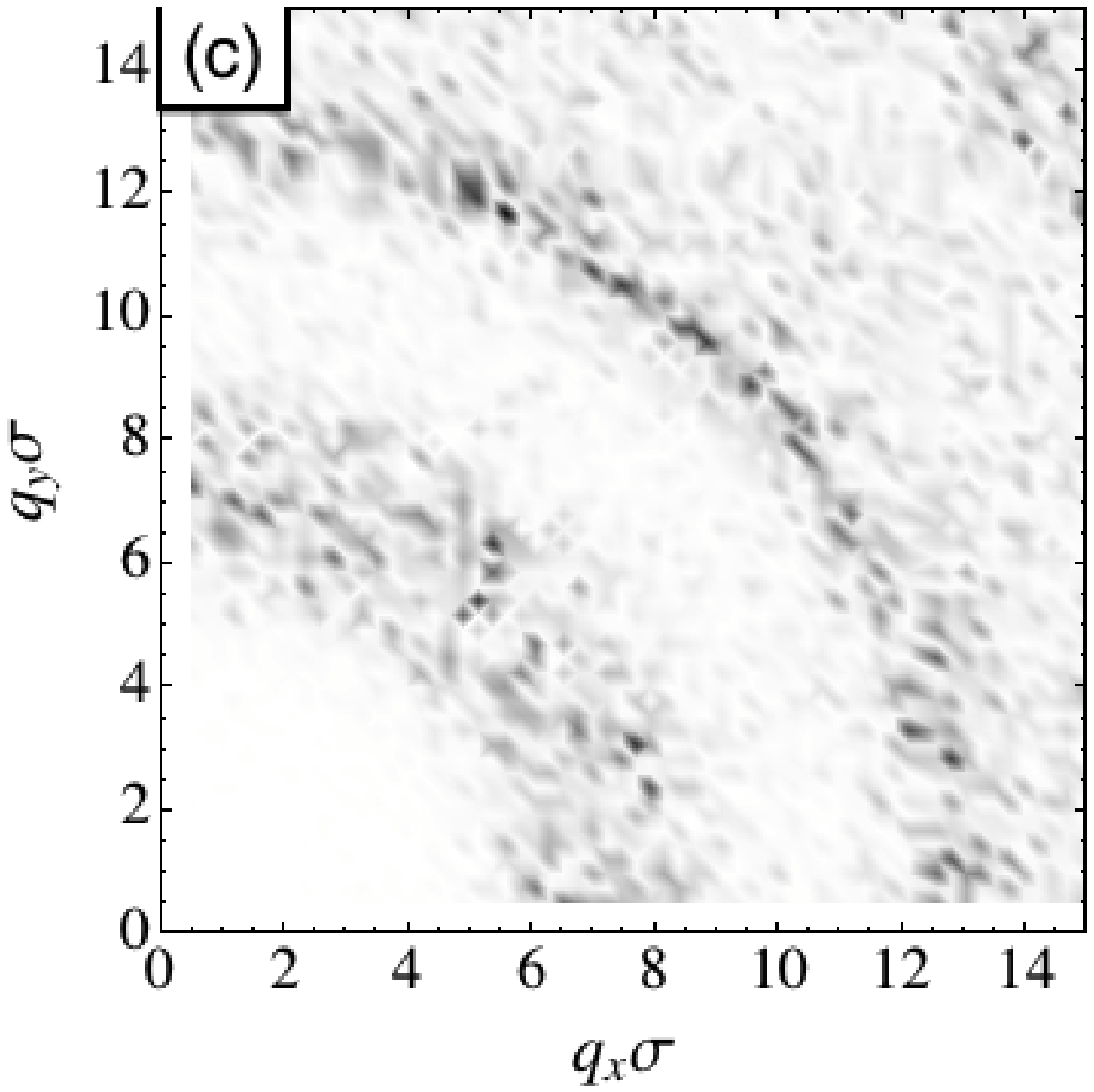}
}
\subfigure{
\label{sqnum6}
\centering\includegraphics[clip=true, trim=0 0 0 0, height=4.0cm, width=4.0cm]{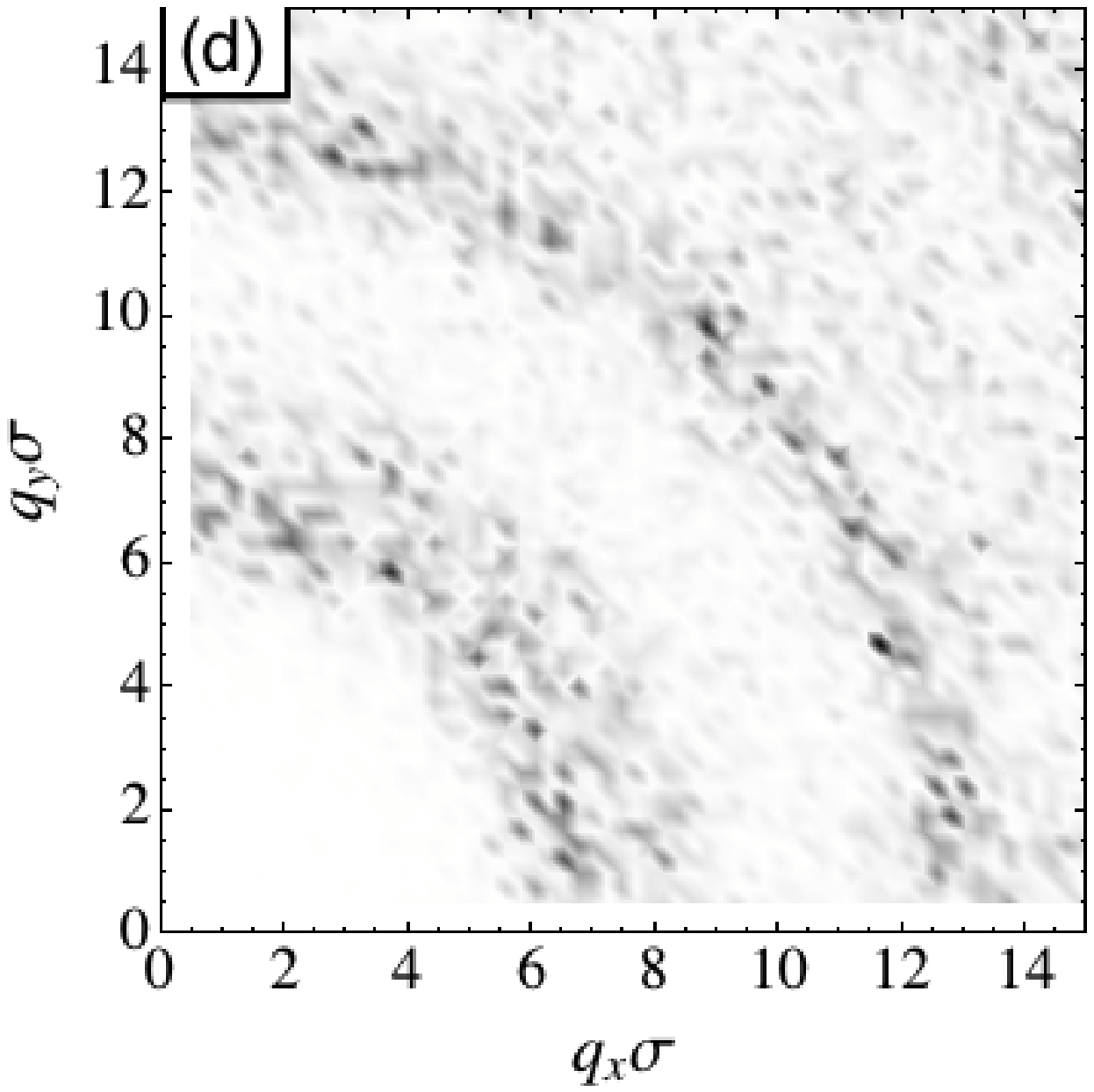}
}
\caption{Structure factor for the $N=65536$ system at $P\sigma^2/\epsilon=5.6$ for the 
(a) S phase at $k_B T/\epsilon=0.30$, 
(b) S at $k_B T/\epsilon=0.40$, 
(c) L at $k_B T/\epsilon=0.29$ and 
(d) L at $k_B T/\epsilon=0.35$. 
The grey scale indicates the value of $S(\vec{q})$ and saturates at a value of 25, which is just above the largest
value for the liquid (large peaks for the S phase appear as white spots with black edges).
}
\label{sofq}
\end{figure}

As an additional measure of order we plot the structure factor in  
Fig.~\ref{sofq} for the same state points considered in Fig.~\ref{g4gg}.
We use a single configuration for the calculation of $S(\vec{q})$, i.e., we do not average over many configurations, 
in order to avoid possible complications arising from rotations of crystal-like domains in time.
The $S(\vec{q})$ for S [panels (a) and (b)] show peaks characteristic of a square crystal.
While there are small hints of scattering for $\vec{q}$ in between the main points located at multiplies
of $\sim 2 \pi/\sigma$ in either $q_x$ or $q_y$, the effect is rather weak compared to what is seen
in other studies of the hexatic phase~\cite{brodin}.
For the liquid, panels (c) and (d) show no hint of crystal-like peaks that might have appeared were 
there a hexatic phase, neither do they show features consistent with a quasi-crystal phase~\cite{skibinsky}.

\section{Discussion and conclusions}

In this study, we vary the parameters $\epsilon_1$, $b$ and $c$ of the SSSW potential
and find that increasing $b$ (the extent of the shoulder)  has the greatest impact on increasing the range of $P$ over which
inverse melting takes place.
Recalculating the melting curve for several values of $b$, we find that the stability field of the S phase
shrinks as a whole while making the effect of inverse melting more pronounced.

For the $b=1.46\sigma$ case, we confirm the melting line predicted by the combination
of several MC free energy methods now becoming standard in the calculation of
phase diagrams by carrying out biased simulations of a phase-separated system.
From these simulations, we estimate the interfacial tension at the inverse melting line 
($P\sigma^2/\epsilon=5.6$, $k_B T/\epsilon=0.340$) 
to be $\beta \gamma \sigma= 0.075$.  This value is rather low compared to 
crystallization in three dimensions, e.g., $\beta \gamma \sigma^2 =0.5$ for hard spheres~\cite{auer},
but is consistent with the rather diffuse interface at coexistence.  A small surface tension is also consistent
with our earlier observations of a small range of metastability of the liquid with respect to crystallization in 
general for this model despite only a small difference in chemical potential at the edge of metastability~\cite{ahmad2}.

The large region of inverse melting for  $b=1.46\sigma$ facilitates direct testing by EDMD simulations.  For both
large and small systems, we confirm the first-order nature of the transition as well as the general location of the transition.

Using the large systems, we test for the range of order.  The orientational correlation function as well as $g( r)$ clearly find the S phase
to be a crystal and L phase to be a liquid.  No hexatic phase is apparent at the point along the melting line where we carry out our 
analyses.  The structure factor likewise supports these findings.  This is consistent with recent work on a simpler 
repulsive-shoulder model in 2D that
has a similar phase diagram to ours, and finds a hexatic phase only at low density~\cite{dudalov}.
Additionally, the structure factor indicates the absence of a quasicrystal phase. 

As for the translational correlation function [$G_{\vec{g}}(r )$], it decays as a power law with a small exponent for S
as is expected.
For the liquid, some care must be taken before exponential decay is made apparent.  First, the orientation of the 
local environment of each origin used in averaging should be taken into consideration when defining lattice
vectors.  Second, one should take into account that the analytical expression for $G_{\vec{g}}(r )$ in the case 
where orientations are uniformly distributed is a Bessel function, for which oscillations decay in amplitude as a power law. 
Thus a spurious power-law decay in translational correlation arises in a completely random system.

Inverse melting in this system, because of the simplicity of the radial pair potential, hopefully will lend
itself to analytical treatment~\cite{urbic}. A more theoretical analysis might be beneficial to understanding 
the impact of other modifications to the potential on inverse melting, and may thus facilitate 
producing inverse melting in other 2D systems that are governed by similar potentials, 
such as lipid membranes~\cite{nielsen}.

\section*{Acknowledgments}

AMA and IS-V thank NSERC for funding, ACEnet for funding and computational support and CFI for funding of
computing infrastructure. SVB acknowledges the partial support of
this research through the Dr. Bernard W. Gamson Computational Science Center at Yeshiva College.


\begin{thebibliography}{999}

\bibitem{stillinger2} F. H. Stillinger and P. G. Debenedetti,
Biophys. Chem. {\bf 105}, 211 (2003).

\bibitem{angiolettiuberti} S. Angioletti-Uberti, B. M. Mognetti, and D. Frenkel,
Nat. Mater. {\bf 11}, 518 (2012).

\bibitem{gang} O. Gang,
Nat. Mater. {\bf 11}, 487 (2012).

\bibitem{dobbs} E. R. Dobbs, 
Helium Three, 
Oxford University Press, Oxford, 2000.

\bibitem{pair} C. Le Pair, K. W. Taconis, R. De Bruyn Ouboter, and P. Das,
Physica {\bf 29}, 755 (1963). 

\bibitem{wilks} J. Wilks, 
The Properties of Liquid and Solid Helium,
Clarendon Press, Oxford, 1967.

\bibitem{rastogi1} S. Rastogi, M. Newman, and A. Keller, 
Nature {\bf 353}, 55 (1991).

\bibitem{rastogi2} S. Rastogi, M. Newman, and A. Keller, 
J. Polym. Sci. B {\bf 31}, 125 (1993).

\bibitem{rastogi3} S. Rastogi, G. W. H. H{\"o}hne, and A. Keller, 
Macromolecules {\bf 32}, 8897 (1999).

\bibitem{greer} A. L. Greer, 
Nature {\bf 404}, 134 (2000).

\bibitem{corstjens} C. S. J. van Hooy-Corstjens, G. W. H. H{\"o}hne, and S. Rastogi,
Macromolecules {\bf 38}, 1814 (2005).

\bibitem{feeney} M. R. Feeney, P. G. Debenedetti, and F. H. Stillinger,
J. Chem. Phys. {\bf 119}, 4582 (2003).

\bibitem{schupper} N. Schupper and N. M. Shnerb,
Phys. Rev. Lett. {\bf 93}, 037202 (2004).

\bibitem{cladis} P. E. Cladis, D. Guillon, F. R. Bouchet, and P. L. Finn, 
Phys. Rev. A {\bf 23}, 2594 (1981).

\bibitem{mortensen} K. Mortensen, W. Brown, and B. Nord{\'e}n,
Phys. Rev. Lett. {\bf 68}, 2340 (1992).

\bibitem{plazanet} M. Plazanet, C. Floare, M. R. Johnson, R. Schweins, and H. P. Trommsdorff,
J. Chem. Phys. {\bf 121}, 5031 (2004).

\bibitem{angelini1} R. Angelini and G. Ruocco,
Philos. Mag. {\bf 87}, 553 (2007).

\bibitem{angelini2} R. Angelini, G. Ruocco, and S. De Panfilis,
Phys. Rev. E {\bf 78}, 020502(R) (2008).

\bibitem{avraham} N. Avraham, B. Khaykovich, Y. Myasoedov, M. Rappaport, H. Shtrickman, D. E. Feldman, T. Tamegai, P. H. Kes, M. Li, M. Konczykowski, K. van der Beek, and E. Zeldov,
Nature {\bf 411}, 451 (2001).

\bibitem{beidenkopf} H. Beidenkopf, N. Avraham, Y. Myasoedov, H. Shtrikman, E. Zeldov, B. Rosenstein, E. H. Brandt, and T. Tamegai,
Phys. Rev. Lett. {\bf 95}, 257004 (2005).

\bibitem{kauzmann} W. Kauzmann,
Chem. Rev. {\bf 43}, 219 (1948).

\bibitem{tombari} E. Tombari, C. Ferrari, G. Salvetti, and G. P. Johari,
J. Chem. Phys. {\bf 123}, 051104 (2005).

\bibitem{vargas} S. Rold{\'a}n-Vargas, F. Smallenburg, W. Kob, and F. Sciortino,
Sci. Rep. {\bf 3}, 2451 (2013).

\bibitem{ahmad2} A. M. Almudallal, S. V. Buldyrev, and I. Saika-Voivod, 
J. Chem. Phys. {\bf 137}, 034507 (2012).

\bibitem{skibinsky} A. Skibinsky, S. V. Buldyrev, A. Scala, S. Havlin, and H. E. Stanley,
Phys. Rev. E {\bf 60}, 2664 (1999).

\bibitem{hemmer} P. C. Hemmer and G. Stell, 
Phys. Rev. Lett. {\bf 24}, 1284 (1970).

\bibitem{stell} G. Stell and P. C. Hemmer, 
J. Chem. Phys. {\bf 56}, 4274 (1972).

\bibitem{denton1}A. R. Denton and H. L\"owen,
J. Phys.: Condens. Matter {\bf 9}, L1 (1997).

\bibitem{denton2}A. R. Denton and H. L\"owen,
J. Phys.: Condens. Matter {\bf 9}, 8907 (1997).

\bibitem{mon1} K. K. Mon, N. W. Ashcroft, and G. V. Chester, 
Phys. Rev. B {\bf 19}, 5103 (1979).

\bibitem{selbert} M. Selbert and W. H. Young, 
Phys. Lett. A {\bf 58}, 469 (1976).

\bibitem{levesque} D. Levesque and J. J. Weis, 
Phys. Lett. A {\bf 60}, 473 (1977).

\bibitem{kincaid} J. M. Kincaid and G. Stell, 
Phys. Lett. A {\bf 65}, 131 (1978).

\bibitem{cummings} P. T. Cummings and G. Stell, 
Mol. Phys. {\bf 43}, 1267 (1981).

\bibitem{velasco} E. Velasco, L. Mederos, G. Navascu{\'e}s, P. C. Hemmer, and G. Stell, 
Phys. Rev. Lett. {\bf 85}, 122 (2000).

\bibitem{voronel} A. Voronel, I. paperno, S. Rabinovich, and E. Lapina, 
Phys. Rev. Lett. {\bf 50}, 247 (1983).

\bibitem{kumar} P. Kumar, S. Han, and H. E. Stanley,
J. Phys.: Condens. Matter {\bf 21} 504108 (2009).

\bibitem{debenedetti} P. G. Debenedetti and H. E. Stanley,
Phys. Today {\bf 56}, 40 (2003).

\bibitem{prielmeier} F. X. Prielmeier, E. W. Lang, R. J. Speedy, and H. D. L{\"u}demann, 
Phys. Rev. Lett. {\bf 59}, 1128 (1987). 

\bibitem{kell} G. S. Kell, 
J. Chem. Eng. Data {\bf 20}, 97 (1975).

\bibitem{poole} P. H. Poole, F. Sciortino, U. Essmann, and H. E. Stanley, 
Nature {\bf 360}, 324 (1992).

\bibitem{cho} C. H. Cho, S. Singh,  and G. W. Robinson, 
Phys. Rev. Lett. {\bf 76}, 1651 (1996).

\bibitem{sadr1} M. R. Sadr-Lahijany, A. Scala, S. V. Buldyrev, and H. E. Stanley, 
Phys. Rev. Lett. {\bf 81}, 4895 (1998).

\bibitem{sadr2} M. R. Sadr-Lahijany, A. Scala, S. V. Buldyrev, and H. E. Stanley, 
Phys. Rev. E {\bf 60}, 6714 (1999).

\bibitem{jagla1} E. A. Jagla, 
J. Chem. Phys. {\bf 111}, 8980 (1999).

\bibitem{jagla2} E. A. Jagla, 
Phys. Rev. E {\bf 63}, 061501 (2001).

\bibitem{jagla3} E. A. Jagla, 
Phys. Rev. E {\bf 63}, 061509 (2001).

\bibitem{sadr3} A. Scala, M. R. Sadr-Lahijany, N. Giovambattista, S. V. Buldyrev, and H. E. Stanley, 
Phys. Rev. E {\bf 63}, 041202 (2001).

\bibitem{sergey} S. V. Buldyrev, G. Franzese, N. Giovambattista, G. Malescio, M. R. Sadr-Lahijany, A. Scala, A. Skibinsky, and H. E. Stanley, 
Physica A {\bf 304}, 23 (2002).

\bibitem{saija} S. Prestipino, F. Saija, and P. V. Giaquinta,
J. Chem. Phys. {\bf 137}, 104503 (2012).

\bibitem{franzese} G. Franzese, G. Malescio, A. Skibinsky, S. V. Buldyrev, and H. E. Stanley, 
Nature {\bf 409}, 692 (2001).

\bibitem{mausbach} P. Mausbacha and H. O. May,
Fluid Phase Equilibr {\bf 214}, 1 (2003).

\bibitem{quigley} D Quigley and M. I. J. Probert,
Phys. Rev. E {\bf 71}, 065701(R) (2005).

\bibitem{head1} T. Head-Gordon and F. H. Stillinger, 
J. Chem. Phys. {\bf 98}, 3313 (1993).

\bibitem{head2} F. H. Stillinger and T. Head-Gordon, 
Phys. Rev. E {\bf 47}, 2484 (1993). 

\bibitem{hus} M. Hu{\v s} and T. Urbic,
J. Chem. Phys. {\bf 139}, 114504 (2013).

\bibitem{frenkel1} D. Frenkel and Berend Smit, 
{\it Understanding Molecular Simulation: From algorithms to Applications}, 
San Diego, Academic Press, 2002.

\bibitem{at} M. P. Allen and D. J. Tildesley, 
Computer Simulation of Liquids, 
Oxford University Press, New York, 1989.

\bibitem{rapaport} D. C. Rapaport, 
The Art of Molecular Dynamic Simulation, 
Cambridge University Press, Cambridge, 1995.

\bibitem{adler} B. J. Adler and T. E. Wainwright, 
J. Chem. Phys. {\bf 31}, 459 (1959).

\bibitem{lubachevsky} B. D. Lubachevsky, 
J. Comput. Phys. {\bf 94}, 255 (1991).

\bibitem{frenkel2} D. Frenkel and A. J. C. Ladd, 
J. Chem. Phys. {\bf 81}, 3188 (1984).

\bibitem{barboy} B. Barboy and W. M. Gelbart, 
J. Chem. Phys. {\bf 71}, 3053  (1979).

\bibitem{noro} M. G. Noro and D. Frenkel, 
J. Chem. Phys. {\bf 114}, 2477 (2001).

\bibitem{henderson} D. Henderson, 
Mol. Phys. {\bf 30}, 971 (1975).

\bibitem{boublik} T. Boubl{\'i}k,
Mol. Phys. {\bf 109}, 1575 (2011).

\bibitem{kofke1} D. A. Kofke, 
Mol. Phys. {\bf 78}, 1331 (1993).

\bibitem{kofke2} D. A. Kofke, 
J. Chem. Phys. {\bf 98}, 4149 (1993).

\bibitem{vega2} C. Vega, E. Sanz, J. L. F. Abascal, and E. G. Noya, 
J. Phys.: Condens. Matter, {\bf 20}, 153101 (2008).

\bibitem{romano} F. Romano, E. Sanz, and F. Sciortino,
J. Chem. Phys, {\bf 132}, 184501 (2010).

\bibitem{singer} S. J. Singer and R. Mumaugh,
J. Chem. Phys. {\bf 93}, 1278 (1990).

\bibitem{shirts} M. R. Shirts and J. D. Chodera,
J. Chem. Phys. {\bf 129}, 124105 (2008). We use the "pymbar-2.0beta" implementation
of the MBAR method available from https://simtk.org/home/pymbar.

\bibitem{steinhardt} P. J. Steinhardt, D. R. Nelson, and M. Ronchetti, 
Phys. Rev. B {\bf 28}, 784 (1983).

\bibitem{gasser} U. Gasser,
J. Phys.: Condens. Matter {\bf 21} 203101 (2009).

\bibitem{bernard} E. P. Bernard and W. Krauth,
Phys. Rev. Lett. {\bf 107}, 155704 (2011).

\bibitem{brodin} A. Brodin, A. Nych, U. Ognysta, B. Lev, V. Nazarenko, M. Skarabot, and I. Musevic,
Condens. Matter Phys. {\bf 13}, 33601 (2010).

\bibitem{auer} S. Auer and D. Frenkel,
J. Chem. Phys. {\bf 120}, 3015 (2004).

\bibitem{dudalov}
D.E. Dudalov, Yu.D. Fomin, E.N. Tsiok, and V.N. Ryzhov,
arXiv:1311.7534v1 (2013).

\bibitem{urbic} T. Urbic,
J. Chem. Phys. {\bf 139}, 164515 (2013).

\bibitem{nielsen}
M. Nielsen, L. Miao, J. H. Ipsen, M. J. Zuckermann, and O. G. Mouritsen,
Phys. Rev. E {\bf 59}, 5790 (1999).


\end{thebibliography}
\end{document}